\documentclass[11pt,a4paper]{article}
\usepackage{jheppub}
\usepackage{bm}
\usepackage{multirow}

\newcommand{\mathsym}[1]{{}}

\newcommand{\baz}{\begin{array}{cc}}
\newcommand{\bad}{\begin{array}{ccc}}
\newcommand{\ba}{\begin{array}{c}}
\newcommand{\ea}{\end{array}}
\newcommand{\be}{\begin{equation}}
\newcommand{\ee}{\end{equation}}
\newcommand{\bea}{\begin{eqnarray}}
\newcommand{\eea}{\end{eqnarray}}

\newcommand{\bi}{\begin{itemize}}
\newcommand{\ei}{\end{itemize}}
\newcommand{\bmt}{\begin{pmatrix}}
\newcommand{\emt}{\end{pmatrix}}
\newcommand{\bt}{\begin{tabular}}
\newcommand{\et}{\end{tabular}}

\newcommand{\benu}{\begin{enumerate}}
\newcommand{\eenu}{\end{enumerate}}

\newcommand{\bav}{\begin{array}{cccc}}

\def\gs{\mathrel{
   \rlap{\raise 0.511ex \hbox{$>$}}{\lower 0.511ex \hbox{$\sim$}}}}
\def\ls{\mathrel{
\rlap{\raise 0.511ex \hbox{$<$}}{\lower 0.511ex \hbox{$\sim$}}}}

\numberwithin{equation}{section}

\title{Proton decay and new contribution to $0\nu 2\beta$ decay in SO(10)
 with low-mass $Z^{\prime}$ boson, observable  ~$n-{\bar n}$ oscillation, 
 lepton flavor violation,~and rare kaon decay}
\author[a]{M. K. Parida,}
\author[b]{Ram Lal Awasthi}
\author[c]{and P. K. Sahu}
\affiliation[a]{Center of Excellence in Theoretical. and Mathematical Sciences,\\
Siksha `O' Anusandhan University, Bhubaneswar-751030, India}
\affiliation[b]{Harish-Chandra Research Institute, Chhatnag Road, 
Jhusi, Allahabad-211019, India}
\affiliation[c]{Institute of Physics, Sachivalaya Marg, Bhubaneswar,
    Odisha-751005, India}
\emailAdd{parida.minaketan@gmail.com}
\emailAdd{ramlal@hri.res.in}
\emailAdd{pradip@iopb.res.in}

\abstract{
In the conventional approach to observable $n-{\bar n}$ oscillation 
through Pati-Salam intermediate gauge symmetry in $SO(10)$, 
the canonical seesaw mechanism is also constrained by the symmetry 
breaking scale $M_R \sim M_C \le 10^6$ GeV which yields light 
neutrino masses several orders larger than the neutrino oscillation 
data. Recently this difficulty has been evaded through 
TeV scale gauged inverse seesaw mechanism while predicting 
experimentally verifiable $W_R^{\pm}, Z_R$ bosons with a new dominant 
contribution to $W_L-W_L$ mediated neutrinoless double beta decay and 
other observable phenomena, but with proton lifetime 
far beyond the accessible limits in foreseeable future. 
In the present work, adopting the view that  we may have only a 
nonstandard TeV scale $Z^{\prime}$ gauge boson accessible to the 
Large Hadron Collider while $W_R^{\pm}$ may be heavy and 
currently inaccessible, we show how a class of non-supersymmetric 
$SO(10)$ models allow experimentally verifiable proton 
lifetime and the new contributions to neutrinoless double beta decay 
in the $W_L-W_L$ channel, lepton flavor violating 
branching ratios, observable $n-{\bar n}$ oscillation, and lepto-quark 
gauge boson mediated rare kaon decays. The occurrence 
of Pati-Salam gauge symmetry with unbroken D-parity and 
two gauge couplings at the highest intermediate scale guarantees 
precision unification in such models. This symmetry also ensures
vanishing GUT threshold uncertainy on $\sin^2\theta_W$ or on the highest
intermediate scale. Although the proton lifetime prediction is brought
closer to the ongoing search limits with GUT threshold
effects in the minimal model, no such threshold effects are needed in
a non-minimal model even for lifetimes close to the Super-Kamiokande limit.
 We derive a new analytic expression for the $0\nu\beta\beta$ decay- 
half-life and show how the existing experimental limits impose the 
lower bound on the lightest of the three heavy sterile neutrino masses, 
$M_{S_1}\ge 14\pm 4$ GeV, irrespective of the nature of hierarchy of
light neutrino masses. We also derive a new lower bound on the lepto-quark 
gauge boson mass mediating rare kaon decay, $M_{\rm lepto} \ge
(1.53{\pm 0.06})\times 10^6$ GeV, which is easily accommodated in
the model. The $n-{\bar n}$ mixing times are predicted in the range
$\tau_{n-{\bar n}}\simeq 10^8-10^{13}$~sec covering those accessible to ongoing experiments.}

\begin{document}
\maketitle

\section{\small INTRODUCTION}\label{sec1}
Although the standard model (SM) of strong, weak, and electromagnetic 
interactions has unraveled the gauge origin of fundamental forces 
and the structure of the universe while successfully confronting  
numerous experimental tests, it has a number of limitations. Experimental 
evidences of tiny neutrino masses compared to their charged lepton 
counterparts also raises the fundamental issue on the origin of these 
masses as well as the nature of neutrinos: whether Dirac\cite{Dirac:1925} 
or Majorana\cite{Majorana:1937vz}- a question whose answer rests 
on the detection and confirmation of $0\nu 2\beta$ decay process 
on which there are a number of ongoing experiments \cite{KlapdorKleingrothaus:2000sn,
Aalseth:2002rf,KlapdorKleingrothaus:2004wj,KlapdorKleingrothaus:2006bd, 
Arnaboldi:2008ds,Auger:2012ar,Gando:2012zm,Agostini:2013mzu}. 
The SM predicts lepton flavor violating (LFV) decays many orders smaller 
than the current experimental limits which appear to be compatible 
via supersymmetric theories. Thus, in the absence of supersymmetry so 
far, it is important to explore non-supersymmetric (non-SUSY) 
extensions of the SM with sizable LFV decay branching ratios. 
  
Several limitations of the SM are removed when it is embedded in a 
popular  grand unified theory (GUT) like $SO(10)$ which has potentialities 
to achieve precision unification of the three forces, accommodate tiny 
neutrino masses through various seesaw mechanisms 
\cite{Minkowski:1977sc,Yanagida:1979as,Mohapatra:1979ia,Schechter:1980gr,Magg:1980ut,Lazarides:1980nt,Ibarra:2010xw}, provide spontaneous origins of Parity and CP-violations \cite{Mohapatra:1974gc,Mohapatra:1980yp,Pati:1974yy,Senjanovic:1975rk,Mohapatra:1980qe}, and a host of other interesting 
physical phenomena. Even without any additional flavor symmetry, 
the model succeeds in representing all fermion masses of three generations 
while observable baryon number violating processes are generic among 
its predictions of new physics beyond the SM. Apart from proton decay, 
theoretical models have been proposed for experimentally observable 
signature of the other baryon number violating process such as 
$n-{\bar n}$ and $H-{\bar H}$ oscillations, and double proton decay 
through GUTs out of which  $n-{\bar n}$ oscillation has attracted 
considerable interest. While in most of the conventional models  \cite{Mohapatra:1980qe,Moha-Senj:1982,mkp83,mkpPRD83,Mohapatra:JPG}, 
the intermediate breaking of Pati-Salam gauge symmetry 
$SU(2)_L\times SU(2)_R \times SU(4)_C (\equiv G_{224})$ has been 
exploited at $\mu=M_C\sim 10^6$ GeV, in a very interesting 
recent development it has been proposed \cite{Babu:2012vc,Babu:PRD,Babu:PRL}.  
 to realize the process 
even if the symmetry breaks near the GUT scale such that the SM gauge
symmetry rules all the way down to the electroweak scale. In this model the 
diquark higgs mediating the oscillation process are tuned to have masses 
at the desired intermediate scale or lower. The model also explains the observed 
baryon asymmetry of the universe by a novel mechanism. Prediction of 
$|\Delta(B-L)| \neq 0$ proton decay mode is another special attractive feature 
of this model. The neutrino masses and mixings in these models are 
governed by high scale canonical or type-II seesaw mechanisms with $B-L$ gauge symmetry breaking occurring near the GUT-scale.      

We consider
worth while to pursue the conventional approach mainly because we are
interested in associating all relevant physical mass scales with the spontaneous
breakings of respective intermediate gauge symmetries. A non-standard extra
$Z^{\prime}$  boson which is under experimental
investigation at LHC energies \cite{ATLAS:extrawz,CMS:extrawz} is
possible by extension of the SM. If such an extension is through
Pati-Salam symmetry at higher scale, the model has the interesting
possibility of accommodating observable $n-{\bar n}$ oscillation
 and rare kaon decays. If Pati-Salam symmetry
in turn emerges from a GUT scenario like $SO(10)$, it provides
interesting possibilities of gauge coupling unification and GUT-scale
representation of all charged fermion masses with the prediction of
Dirac neutrino mass matrix. If one fermion singlet per generation is
added to the $SO(10)$ frame work, it has the interesting possibility of
explaining light neutrino masses and mixings by experimentally
verifiable gauged inverse seesaw
mechanism. Whereas the non-supersymmetric SM as
such predicts negligible contributions to charged lepton flavor
violating (LFV) decays, the TeV scale inverse seesaw mechanism predcts
LFV branching ratios only $4-5$ order smaller than the current
experimental limits. 
Embedding such a mechanism through
heavier right-handed neutrinos provides further interesting
realisation of additional new dominant contributions to neutrino-less
double-beta decay in the $W_L-W_L$ channel through the exchanges of
sterile neutrinos which turn out to be Majorana fermions in the model.           
      
In this work we attempt to revive the conventional approach 
\cite{mkp83,mkpPRD83,Chang:1983fu,Chang:1984wp} but by evading the 
light neutrino mass constraint through inverse seesaw formula gauged 
by the TeV scale symmetry $SU(2)_L\times U(1)_R\times U(1)_{B-L} \times SU(3)_C$ 
manifesting in an extra $Z^{\prime}$ boson which might be detected by 
ongoing search experiments at the Large Hadron Collider, a strategy which 
has been adopted recently in non-supersymmetric (non-SUSY) $SO(10)$
grand unified theory (GUT)\cite{ap:2011aa,Awasthi:2013ff}. Low energy signature of lepto-quark 
gauge bosons is also predicted through rare kaon decay 
$K_L\to \mu {\bar e}$ with branching ratios close to the current
experimental limit \cite{dambrose98}. Once the experimentally testable gauged 
inverse seesaw mechanism is made operative, the model 
is found to predict a number of new physical quantities to be 
verified by ongoing search experiments at low and accelerator 
energies.  They include (i) dominant contribution to $0\nu 2\beta$ 
rate in the $W_L-W_L$ channel due to heavy  sterilele neutrino
exchanges leading to the lower bound on the lightest
sterile neutrino mass $M_{S_1} \ge 14\pm 4$ GeV, (ii) unitarity-violating 
contributions to branching ratios for lepton flavor violating 
(LFV) decays, (iii) leptonic CP-violation due to non-unitarity 
effects, (iv) experimentally verifiable  $|\Delta(B-L)|=0$ proton 
decay modes such as $p\to e^+\pi^0$ (v)lepto-quark gauge-boson  
mediated rare kaon decay with ${\rm Br}.(K_L\to \mu {\bar e}) 
\simeq 10^{-12}$, and (vi) observable $n-{\bar n}$-oscillation
mixing time $10^8-10^{13}$ sec with the possibility of a diquark Higgs
scalar at the TeV scale. 

The quark-lepton symmetric origin of the Dirac neutrino mass matrix 
($M_D$) is  found to play a crucial role in enhancing the effective 
mass parameter for $0\nu\beta\beta$ decay. We also briefly discuss 
how a constrained (unconstrained) value of the RH neutrino mass matrix 
emerges from the SO(10) structure with one $126$ (two $126$'s) from the 
GUT-scale fit to charged fermion masses.

Although some of the results of the present work were also derived  
in a recent work \cite{Awasthi:2013ff}, the model  
required the asymmetric left-right gauge symmetry at 
$\simeq 10$ TeV leading to the prediction of $W_R^{\pm}, Z_R$ gauge bosons 
at LHC energies. In the present $SO(10)$ model, we show that even 
though  only a TeV scale $Z^{\prime}$ boson \cite{zprime-th} is detected at the  LHC, 
a number of these observable predictions are still 
applicable even if the $W_R^{\pm}$ boson masses are beyond the 
currently accessible LHC limit. In contrast to the earlier 
model, in the present work we predict proton lifetime to be accessible 
to ongoing search experiments. The symmetry breaking chain of the model 
is found to require $SU(2)_L\times SU(2)_R\times SU(4)_C\times 
D (g_{2L}=g_{2R})(\equiv g_{224D})$ gauge symmetry at the highest 
intermediate scale ($M_P$) which eliminates the possible presence of triangular 
geometry of gauge couplings around the GUT scale. This in turn determines 
the unification mass precisely,  
at the meeting point of two gauge coupling constant lines. 
The other advantage of this symmetry is that it pushes most of the larger- 
sized submultiplets down to the parity restoring intermediate scale reducing 
the size of GUT-threshold effects on the unification scale 
and proton lifetime while the GUT-threshold effects on $\sin^2\theta_W$ 
or $M_P$ have exactly vanishing contribution \cite{parpat:1991,parpat:1992,mkp:1998}. 

We derive a new formula for half-life of $0\nu\beta\beta$ decay as 
function  of three heavy sterile neutrino masses and  
 provide a new  plot of the half-life against the lightest of the
 three masses. From   the existing experimental values of life-time
 \cite{KlapdorKleingrothaus:2006bd,Auger:2012ar,Gando:2012zm,Agostini:2013mzu},
 we deive the lower limit on the sterile neutrino mass of the first 
generation ${ M_{S_1}> 14\pm 4 }$ GeV which is also 
experimentally verifiable. This result is found to hold irrespective
of the nature of hierarchy of light neutrino masses. We find that the
saturation of half-life obtained by Heidelberg-Moscow experiment does
not necessarily require light neutrinos to be quasi-degenerate. These
results on the half-life prediction on $0\nu\beta\beta$ decay and the
derived bound on $M_{S_1}$ is also
applicable to the model of ref.\cite{Awasthi:2013ff}.

This paper is organized 
in the following manner. In Sec.~2 we discuss the 
specific $SO(10)$ symmetry breaking chain and study predictions 
of different physically relevant mass scales emerging as solutions 
to renormalization group equations. In Sec.~3 we discuss predictions of 
proton lifetime accessible to ongoing search experiments. Lower bound
on the lepto-quark gauge boson mediating rare-kaon decay is derived in
Sec.4 where mixing times for $n-{\bar n}$ oscillation are also
predicted.
 In Sec.~5 we 
discuss the derivation of Dirac neutrino mass matrix from GUT-scale 
fit to the charged fermion masses where, in a minimal $SO(10)$ 
structure, we also show how the model predicts the RH neutrino 
mass eigenvalues which can be detected at the LHC. Fits to the 
neutrino oscillation data are discussed in Sec.~6. In Sec.~7 
we discuss the model estimations of LFV decay branching ratios 
and CP-violating parameter due to non-unitarity effects. In Sec.~8 
we obtain the model estimations on the dominant contributions 
to $0\nu 2\beta$ process and study variation of half-life as a 
function of sterile neutrino masses. The paper is summarized 
with conclusions in Sec.~9. In the Appendix we derive analytic 
formulas for GUT threshold effects on ${\rm ln}(M_P/M_Z)$ 
and ${\rm ln}(M_U/M_Z)$.             
\section{\small PRECISION GAUGE COUPLING UNIFICATION AND MASS SCALES}\label{sec2}
In the conventional approach to investigation of gauge coupling 
unification, usually the semi simple gauge symmetry to which the 
GUT gauge theory breaks is a product of three or four individual 
groups. As a result the symmetry below the GUT-breaking scale 
involves three or four gauge couplings. The most popular of 
such examples is $SO(10)\to G_I$ where $G_I=SU(2)_L\times U(1)_Y\times SU(3)_C$ 
which has three different gauge couplings, $g_Y$, $g_{2L}$ and 
$g_{3C}$, whose renormalization group (RG) evolution creates a triangular 
region around the projected unification scale making the determination
of the scale more or 
less uncertain. Even though the region of uncertainty is reduced 
in the presence of intermediate scales, it exists in principle when, 
for example, $G_I=SU(2)_L\times SU(2)_R\times U(1)_{B-L}\times 
SU(3)_C$, that included three or four gauge couplings. 
Only in the case when $G_I=SU(2)_L\times SU(2)_R\times 
SU(4)_C\times D$, the Pati-Salam symmetry with LR discrete symmetry 
\cite{Pati:1974yy} ($\equiv$ D-Parity) \cite{Chang:1983fu,Chang:1984wp}, 
there are two gauge couplings $g_{2L}=g_{2R}$ and $g_{4C}$, and 
the meeting point of the two RG-evolved coupling lines determines 
the unification point exactly. Several interesting consequences 
of this intermediate symmetry have been derived earlier including 
vanishing corrections to GUT-threshold effects on $\sin^2{\theta_W}$ 
and the intermediate scale \cite{parpat:1991,parpat:1992,mkp:1998,rnmmkp:1992}. 
We find this symmetry to be essentially required at the highest 
intermediate scale in the present model to guarantee several 
observable phenomena as $SO(10)$ model predictions while safeguarding 
precision unification.       
 
We consider the symmetry breaking chain of non-SUSY SO(10) GUT 
which gives a rich structure of new physics beyond the SM provided 
the Pati-Salam symmetry occurs as an intermediate symmetry twice: 
once between the  high parity breaking scale ($M_P$) and the GUT 
scale ($M_{\rm GUT}$) and, for the second time, without parity 
between the $SU(4)_C$ breaking scale ($M_C$) and $M_P$
\begin{eqnarray}
SO(10) & \stackrel{M_U}{\rightarrow} & SU(2)_L \times SU(2)_R 
\times SU(4)_C \times D \quad [{G}_{224D},\, g_{2L} = g_{2R}]\nonumber \\
       & \stackrel{M_P}{\rightarrow} & SU(2)_L \times SU(2)_R \times 
       SU(4)_C \quad [{G}_{224},\, g_{2L} \neq g_{2R}] \nonumber \\
       & \stackrel{M_C}{\rightarrow} & SU(2)_L \times U(1)_R \times 
       U(1)_{\rm (B-L)} \times SU(3)_C \quad [{G}_{2113}] \nonumber \\
       & \stackrel{M_R}{\rightarrow} & SU(2)_L \times U(1)_Y \times SU(3)_C  
       \quad [{G}_{\rm SM}] \nonumber \\
       &\stackrel{m_W}{\rightarrow}  & U(1)_{\rm em} 
       \times SU(3)_C \quad [{G}_{13}]\,.  \nonumber
\end{eqnarray}

The first step of spontaneous symmetry breaking is implemented 
by giving GUT scale VEV to the D-Parity even Pati-Salam 
singlet contained in ${54}_H\subset SO(10)$ leading to the 
left-right symmetric gauge group $G_{224D}$ with the equality 
in the corresponding gauge couplings $g_{2L}=g_{2R}$. The 
second step of breaking occurs by assigning Parity breaking VEV 
to the D-Parity odd singlet $\eta(1,1,1)\subset {210}_H$ 
\cite{Chang:1983fu,Chang:1984wp} resulting in the LR asymmetric 
gauge theory $G_{224}(g_{2L}\neq g_{2R})$. The third step of 
breaking to  gauge symmetry $G_{2113}$ is implemented by assigning VEV 
of order $M_C \sim 10^5-10^6$~GeV to the neutral component of 
the $G_{224}$ sub-multiplet $(1,3,15)\subset {210}_H$. This technique 
of symmetry breaking to examine the feasibility of observable 
$n-{\bar n}$ through the type of intermediate breaking $G_{224} \to G_{2113}$ was proposed 
at a time when neither the neutrino oscillation data, nor
the precision CERN-LEP data were available~\cite{mkp83,mkpPRD83,Chang:1984wp}. The gauge 
symmetry $G_{2113}$ that is found to survive down to the TeV scale 
is broken to the SM by the sub-multiplet $\Delta_R(3,1,
{\bar {10}})\subset {126}_H$ leading to the low-mass extra $Z^{\prime}$ 
boson accessible to LHC. At this stage RH Majorana mass matrix 
$M_N=f\,\langle \Delta_R^0\rangle$ is generated through the Higgs Yukawa 
interaction. The last step of breaking occurs as usual through 
the VEV of the SM doublet contained in the sub-multiplet 
$\phi(2,2,1)\subset {10}_H$. 
The VEV of the neutral component of RH Higgs doublet $\xi_R(1,2,4)$ under 
$G_{224}$ symmetry contained in $16_H\,(SO(10)$ is used to generate the 
$N-S$ mixing mass term needed for extended seesaw mechanism.
For the sake of fermion mass 
fit at the GUT, scale we utilize two Higgs doublets for $\mu \ge 
5$ TeV. Out of these two, the up type doublet $\phi_u \subset 
{10}_{H_1}$ contributes to Dirac masses for up quarks and 
neutrinos, and the down type doublet  $\phi_d \subset {10}_{H_2}$  
contributes to masses of down type quarks and charged 
leptons. We will see later in this work how the induced 
VEV of the sub-multiplet $\xi(2,2,15)\subset {126}_H$ \cite{Babu:1992ia,Awasthi:2013ff} 
naturally available in this model plays a crucial role in splitting 
quark and lepton masses at the GUT scale and determining 
the value of $M_D$. In one interesting scenario, the GUT 
scale fit to fermion masses and mixings results in the diagonal 
structure of RH neutrino mass matrix near the TeV scale which 
is accessible for verification at LHC energies. 

Using extended survival hypothesis \cite{Aguila:1981,Moha-Senj:1982} the Higgs scalars responsible 
for spontaneous symmetry breaking and their contributions 
to $\beta-$ function coefficients up to two-loop order are 
given in Tab.\ref{tab:beta_coeff}. One set of allowed solutions for mass scales 
and GUT-scale fine-structure constant is
\bea
 M_R^0&=&5 ~{\rm TeV},  M_{\Delta}=M_C=10^{5.5}-10^{6.5}~ {\rm GeV}, M_P=10^{13.45}~ {\rm GeV},\nonumber\\
 M_{\rm GUT}&=&10^{16.07}~ {\rm GeV},~~\alpha_G= 0.0429. \label{rgsol}
\eea
where $M_{\Delta}$ represents the degenerate mass of diquark Higgs
  scalars contained in
  $\Delta_R(1,3,{\bar {10}})\subset {126}_H$. 

\begin{figure}[htb!]
\centering
\includegraphics[scale=1.0,angle=0]{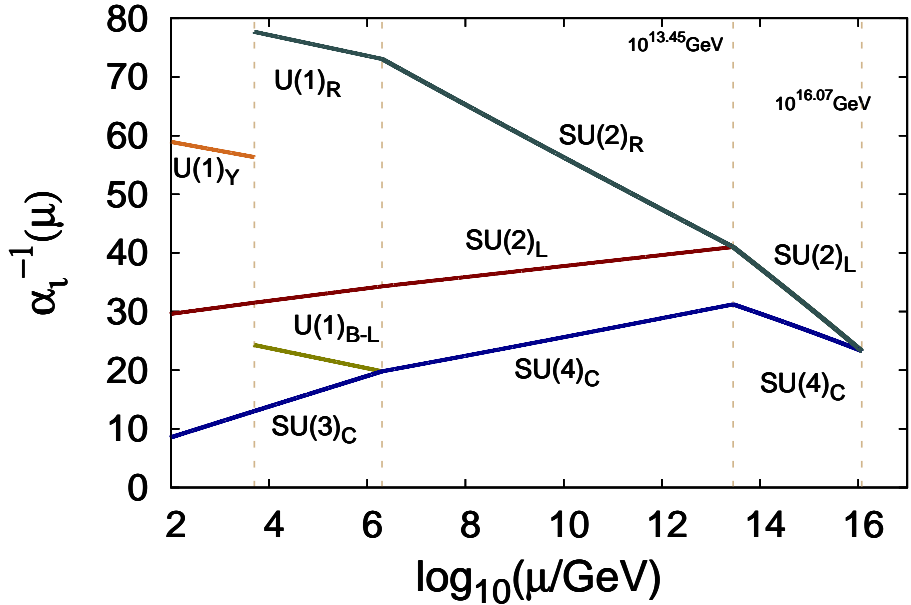}
\caption{gauge coupling unification including $\xi(2,2,15)$}
\label{fig:unify}
\end{figure}

The renormalization group evolution of gauge couplings is 
shown in Fig.\ref{fig:unify}  exhibiting precision unification. 

We have noted that when $M_{\Delta}< M_C$, there is a small decrease
in the unification scale that is capable of reducing the proton lifetime
predictions by a factor $3-5$. One example of this solution is,
\bea
 M_R^0&=&5 ~{\rm TeV},  M_{\Delta}=10^4~{\rm GeV}, M_C=10^{6}~ {\rm GeV}, M_P=10^{12.75}~ {\rm GeV},\nonumber\\
 M_{\rm GUT}&=&10^{15.92}~ {\rm GeV},~~\alpha_G= 0.0429. \label{rgsol-del}
\eea
It is interesting to note that the present LHC bound on the diquark
Higgs scalar mass \cite{CMS:diquark} is
\be
{(M_{\Delta})}_{expt.} \ge 3.75 {\rm TeV}.\label{diquark-LHC}
\ee
As discussed in the context of $n-{\bar n}$ oscillation in Sec.4, our model
accommodates a TeV scale diquark with observable mixing time. 
But substantial decrease in the
unification scale and the corresponding decrease in proton lifetime is
possible when the bi-triplet Higgs scalar 
${\bf\Theta}_H(3,3,1)\subset {54}_H$ is lighter than the GUT scale by
a factor ranging from ${1\over{15}}-{1\over{25}}$. These solutions
are discussed in the following section.

\section{{\small LOW MASS $Z^{\prime}$ AND PROTON DECAY}}\label{sec3}
\subsection{Low-mass  $Z^{\prime}$ boson}
In the solutions of RGEs with precision unification, we have found 
that $g_{(B-L)}=0.72-0.75$ and $g_{1R}=0.40-0.42$ 
in the range of values $M_R^0=v_{B-L}= 3-10$ TeV. This predicts 
the mass of the $Z^{\prime}$ boson in the range  
\be
M_{\rm Z^{\prime}}= 1.75 - 6.1 ~{\rm TeV},\label{zprimemass}  
\ee
whereas the current experimental bound from LHC is  
$ (M_{\rm Z^{\prime}})_{\rm expt.} \ge 2.5$ TeV. Thus, if such a
 $Z^{\prime}$ boson in the predicted mass range of the present model
exists, it is likely to be discovered by the ongoing searches at the LHC. 

\subsection{Proton Lifetime for $p\to e^+\pi^0$}\label{plifesub}   
\subsubsection{Predictions at two-loop level}
The formula for the inverse 
of proton-decay width\cite{Pnath:2007,Parida:2011wh,Bertolini:2013vta} is
\be
\Gamma^{-1}(p\to e^+\pi^0)
 =\frac{64\pi f_{\pi}^2}{m_p}
 \left(\frac{{M_U}^4}{{g_G}^4}\right)\frac{1}{|A_L|^2|\bar{\alpha_H}|^2(1+D+F)^2\times R}.
\label{inversewidth}
\ee
where  $R=[(A_{SR}^2+A_{SL}^2) (1+ |{V_{ud}}|^2)^2]$ for $SO(10)$, $V_{ud}=0.974=$ the  
$(1,1)$ element of $V_{CKM}$ for quark mixings, $A_{SL}(A_{SR})$ is the short-distance 
renormalization factor in the left (right) sectors and $A_L=1.25=$ long distance 
renormalization factor. $M_U=$ degenerate mass of $24$ superheavy gauge bosons in
$SO(10)$, $\bar\alpha_H =$ hadronic matrix element, $m_p = $ proton mass $=938.3$ MeV, 
$f_{\pi}=$ pion decay constant $=139$ MeV, and the chiral Lagrangian parameters are 
$D=0.81$ and $F=0.47$. Here $\alpha_H= \bar{\alpha_H}(1+D+F)=0.012$ GeV$^3$ is obtained
from lattice gauge theory computations. In our model, the product of the short 
distance with the long distance renormalization factor $A_L=1.25$ turns out to 
be  $A_R \simeq A_LA_{SL}\simeq A_LA_{SR}\simeq 3.20$. Then using the the 
two-loop value of the unification scale and the GUT coupling from 
eq.(\ref{rgsol}) gives
\be
{\tau}_p(p\to e^+\pi^0) \simeq 5.05 \times 10^{35} {\rm yrs}\label{plifeth}
\ee
whereas the solution of RGEs corresponding to eq.(\ref{rgsol-del})
gives
\be
{\tau}_p(p\to e^+\pi^0) \simeq 1.05 \times 10^{35} {\rm yrs}.\label{plifeth-del}
\ee
For comparison we note the current experimental search 
limit from Super-Kamiokande 
is~\cite{Nishino:2012ipa,jraaf:2012,Hewett:2012,babuetal:2013} 
\be
(\tau_p)_{\rm Super K.} \ge 1.4 \times 10^{34} {\rm yrs}.\label{superk}
\ee
A second generation underground water cherenkov detector 
being planned at Hyper-Kamiokande in Japan is expected to probe higher 
limits through its $5.6$ Megaton year exposure leading to the 
partial lifetime~\cite{babuetal:2013}
\be
(\tau_p)_{\rm Hyper K.} \ge 1.3 \times 10^{35} {\rm yrs}.\label{hyperk}
\ee 
if actual decay event is not observed within this limit.
Thus our model prediction in eq.(\ref{plifeth-del}) barely within the
 planned Hyper-K limit although this the prediction in eq.(\ref{plifeth})
nearly $4$ times larger than this limit.
 
If the proton decay is observed closer to the current or planned 
experimental limits, it would vindicate the long 
standing fundamental hypothesis of grand unification. On the other
hand proton may be much more stable and its lifetime may not be
accessible even to Hyper K. experimental search programme. These
possibilities are addressed below. \\

\subsubsection{GUT scale and proton life-time reduction through bi-triplet scalar}
\begin{figure}[htb!]
\centering
\includegraphics[scale=1.0,angle=0]{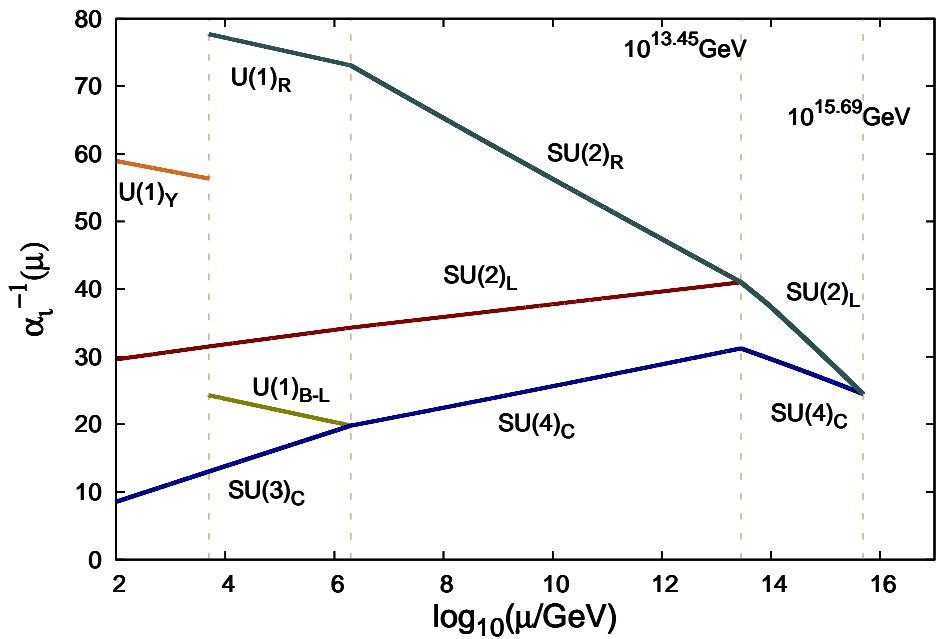}
\caption{Same as Figure 1 but with the Higgs scalar bi-triplet
   of mass $9\times 10^{13}$ GeV.}
\label{fig:unify-2}
\end{figure}

We note that the present estimation of the GUT scale can
be significantly lowered so as to bring the proton-lifetime prediction
 closer to the current Super-K. limit if the the Higgs scalar
 bi-triplet ${\large\bf {\Theta}}_H(3,3,1)\subset {54}_H$ of $SO(10)$ is
 near the Parity violating intermediate scale.
For example in Figure \ref{fig:unify-2}, we have shown how in this model only the
unification scale is lowered while keeping the other physical mass
scales unchanged as in eq.(\ref{rgsol}) for a value of $M_{331}=9\times 10^{13}$
GeV.

In Table \ref{tab:M331-effect} we have presented various allowed
values of the GUT scale and the proton life-time for different
combinations of the diquark Higgs scalar masses $M_{\Delta}$ contained
in $\Delta_R(1,3,{\bar{10}}) \subset {126}_H$ which mediate $n-{\bar
  n}$ oscillation process. Even for a the bi-triplet mass $M_{GUT}/15$
we note a reduced value of the unification scale at $M_U=10^{15.63}$
GeV and the corresoponding proton lifetime at $\tau_P=4.6 \times
10^{33}$ yrs when $M_{\Delta} \sim 10^4$ GeV. The estimated lifetimes
without including the GUT-threshold effects is found to be in the
range $\tau_P=4.6 \times
10^{33}$ yrs to $2.1\times 10^{35}$ yrs, most of which are between the
Super-K and the Hyper-K limits.
\begin{table}
\begin{center}
\begin{tabular}{|l|l|l|l|l|l|}
\hline
$M_{\Delta}$~(GeV)&$M_P$~(GeV)&
$M_{(3,3,1)}$~(GeV) & $M_G$~(GeV)
&$\alpha^{-1}_G$& $\tau_p({\rm years})$ \\
\hline
$10^{4.0}$&$10^{12.73}$&$10^{14.00}$& $10^{15.57}$ & 22.37 & $4.65\times 10^{33}$\\
\hline
$10^{4.0}$&$10^{12.73}$&$10^{14.50}$& $10^{15.66}$ & 22.08 & $1.03\times 10^{34}$\\
\hline
$10^{4.0}$&$10^{12.73}$&$10^{15.00}$& $10^{15.75}$ & 21.79 & $2.32\times 10^{34}$\\
\hline
$10^{4.0}$&$10^{12.73}$&$10^{15.92}$& $10^{15.92}$ & 21.22 & $1.05\times 10^{35}$\\
\hline
$10^{4.5}$&$10^{12.89}$&$10^{14.00}$& $10^{15.60}$ & 23.16 & $6.58\times 10^{33}$\\
\hline
$10^{4.5}$&$10^{12.89}$&$10^{14.50}$& $10^{15.69}$ & 22.88 & $1.47\times 10^{34}$\\
\hline
$10^{4.5}$&$10^{12.89}$&$10^{15.50}$& $10^{15.87}$ & 22.19 & $7.26\times 10^{34}$\\
\hline
$10^{4.5}$&$10^{12.89}$&$10^{15.95}$& $10^{15.95}$ & 22.01 & $1.49\times 10^{35}$\\
\hline
$10^{5.0}$&$10^{13.05}$&$10^{14.00}$& $10^{15.62}$ & 23.94 & $8.45\times 10^{33}$\\
\hline
$10^{5.0}$&$10^{13.05}$&$10^{14.50}$& $10^{15.71}$ & 23.66 & $1.89\times 10^{34}$\\
\hline
$10^{5.0}$&$10^{13.05}$&$10^{15.50}$& $10^{15.89}$ & 23.08 & $9.44\times 10^{34}$\\
\hline
$10^{5.0}$&$10^{13.05}$&$10^{15.98}$& $10^{15.98}$ & 22.79 & $2.11\times 10^{35}$\\
\hline
\end{tabular}
\caption{Predictions on lifetime for the decay $p\to e^+\pi^0$ with
  lower values of masses of the bi-triplet and the diquark Higgs
  scalars.}
\label{tab:M331-effect}
\end{center}
\end{table}

An important source of uncertainty on $\tau_P$ in GUTs is known to be   due to GUT-threshold effects as illustrated
 in the following sub-section.

\subsubsection{Estimation of GUT-threshold effects}
That there could be significant threshold effects on the unification
scale arising out of heavy and super-heavy particle masses was pointed out especially in the context grand desert
models \cite{Weinberg:1980,Hall:1981,Ovrut:1981,Langacker:1993} and in
intermediate scale $SO(10)$ models
\cite{parpat:1991,parpat:1992,mkp:1998,rnmPLB:1993,rnmmkp:1992,lmpr:1994vp,mkp:other-th}. 
 
In order to examine how closer to or farther from the current 
experimental bound our  model predictions could be,  
we have estimated the major source of uncertainty on proton 
lifetime due to GUT threshold effects in $SO(10)$ with intermediate scales \cite{rnmmkp:1992,lmpr:1994vp} 
taking into account the contributions of the superheavy (SH) 
components in ${54}_H,{126}_H,{210}_H, {10}_{H_1}$ and
${10}_{H_2}$ in the case of the minimal model
\begin{eqnarray}
{210}_H &\supset& \Sigma_1 (2,2,10)+\Sigma_2(2,2,{\overline {10}})
+\Sigma_3(2,2,6)+ \Sigma_4 (1,1,15), \nonumber\\
{54}_H&\supset & S_1(1,1,{20}+S_2(3.3,1)+S_3(2,2,6),\nonumber\\ 
{126}_H&\supset& \Delta_1(1,1, 6),~~ {10}_{H_i}\supset H_i(1,1,6), i=1,2,\label{shcomp}
\end{eqnarray}
where the quantum numbers on the RHS are under the gauge group 
$G_{224}$ and the components have superheavy masses around 
the GUT scale. It was shown in refs.\cite{parpat:1991,parpat:1992,mkp:1998} that 
when $G_{224D}$ occurs as intermediate symmetry, all loop 
corrections due to superheavy masses  $m_{SH} \ge M_P$  cancels 
out from the predictions of $\sin^2\theta_W$ and also 
from $M_P$ obtained as solutions of RGEs for gauge couplings 
while the GUT threshold effect on the unification scale due to the superheavy scalar 
masses assumes an analytically  simple form. As outlined in 
the Appendix, even in the presence of two more intermediate 
symmetries below $M_P$, analogous formulas on the GUT-threshold 
effects are also valid 
\be
\Delta {\rm ln}({M_{\rm U}\over M_{\rm Z}})
	=\frac{\lambda_{2L}^U-\lambda_{4C}^U}{6( 
     a'''_{2L}-a'''_{4C})} \label{thformula}
\ee 
where $a'''_i$ is one-loop beta function coefficients in the 
range $\mu= M_P-M_U$ for the gauge group $G_{224D}$.
In eq.(\ref{thformula}) 
\be
\lambda_i^U=b^V_i+{\Sigma}_{\rm SH}b_i^{\rm SH}{\rm ln}
({M_{SH}\over M_{U}}), i=2L, 2R, 4C
\label{lambdadef}
\ee
$b^V_i=tr(\theta^V_i)^2$ and $b^{\rm SH}_i=tr(\theta^{\rm SH}_i)^2$ 
where $\theta^V_i$ ($\theta^{\rm SH}_i$) are generators of the 
gauge group $G_{224D}$ in the representations of superheavy gauge 
bosons (Higgs scalars). The one-loop coefficients for various SH components 
in eq.(\ref{shcomp}) contributing to threshold effects are 
\cite{rnmmkp:1992}
\bea
b^V_{2L}&=&b^V_{2R}= 6,~~b^V_{4C}=4,~b^{\Sigma_4}_i=(0,0,4) \nonumber\\ 
b^{\Sigma_1}_i&=& b^{\Sigma_2}_i=(10,10,12), ~b^{\Sigma_3}_i= b^{S_3}_i=(6,6,4), \nonumber\\
b^{S_1}_i&=&(0,0,16),~b^{S_2}_i= (12,12,0),~b^{H_{1,2}}_i=b^{\Delta_1}_i=(0,0,2), \label{dynkin}
\eea
where we have projected out the would-be Goldstone components from $S_3$ leading to
\be
 \lambda_{2L}^U-\lambda_{4C}^U=2-6\eta_{210}-2\eta_{54}-2\eta_{126}-4\eta_{10},\label{lambdalu}
\ee
with $\eta_X= {\ln}(M_X/M_U)$, and we have made the plausible 
assumption that all SH scalars belonging to a particular 
$SO(10)$ representation have a common mass such as 
$M_{210}=M_{\Sigma_i} (i=1-4)$ for ${210}_H$ and so on for other representations 
\cite{lmpr:1994vp}. Utilising the model coefficients $a'''_{2L}= 44/3$ 
and $a'''_{4C}= 16/3$,  and using eq. (\ref{lambdalu}) in eq.(\ref{thformula}) 
gives
\be
M_U/M_U^0= 10^{(0.25\ln\eta)/2.3025} \label{gutthresexp2}
\ee
where $\eta =10(1/10)$ depending upon our assumption that SH 
components are 10(1/10) times heavier(lighter) than the GUT scale. By
applying these GUT-threshold effects to the solutions of RGE in
eq.(\ref{rgsol-del}), we obtain
\begin{eqnarray} 
M_U&=&10^{15.92\pm 0.25} {\rm GeV}, \nonumber\\
{\tau}_p(p\to e^+\pi^0) &\simeq& 5.05 \times 10^{35\pm 1.0 \pm 0.34 } {\rm yrs}
\label{plifethrs}   
\end{eqnarray}
 where the first uncertainty is due to GUT threshold effects, 
 and the second uncertainty derived in Appendix B, is due 
 to  the $1\sigma$ level uncertainties in the experimental 
 values of $\sin^2\theta_W(M_Z)$ and $\alpha_S(M_Z)$. It is clear 
 from eq.(\ref{plifethrs}) that our prediction covers wider range 
 of values in proton lifetime prediction including value few times
 larger than the current Super-K. limit. 

Similarly each of the numerical
 values in the last column of Table \ref{tab:M331-effect} is modified
 by this additional uncertainty factor of $10^{\pm 1 \pm 0.32}$ in the estimated
 lifetimes.

 \section{\small RARE KAON DECAY AND {\Large $n-{\bar n}$} OSCILLATION}
In this section we discuss the model predictions on rare kaon decays 
mediated by lepto-quark gauge bosons of $SU(4)_C$ that 
occurs as a part of Pati-Salam intermediate gauge symmetry 
$SU(2)_L\times SU(2)_R\times SU(4)_C$ which breaks spontaneously 
at the mass scale $\mu=M_R^+=M_C$. The lepto-quark
Higgs scalar contribution to the rare decay process is suppressed in
this model due
to the natural values of their masses at $M_C = 10^6$ GeV and smaller
Yukawa couplings. \\

\subsection{Rare Kaon Decay $K_L \to \mu {\bar e}$}
Earlier several attempts have been made to derive lower bound 
on the lepto-quark gauge boson mass \cite{Pati:1974yy,Deshpande:1982bf,Parida:1996xa}.
In this subsection we update the existing latest lower bound on 
the $SU(4)_C$-leptoquark gauge boson mass \cite{Parida:1996xa} 
using the improved measurement on the branching ratio and improved 
renormalization group running of gauge couplings due 
to the running VEV's and the additional presence of $G_{2113}$ 
gauge symmetry in between $G_{224}$ and the SM gauge symmetries.  
Experimental searches on rare kaon decays  in the channel 
$K_L \to \mu^{\pm} e^{\mp}$ have limited its  branching ratio 
with the upper bound \cite{dambrose98}
\be
\mbox{Br.}\left(K_L \to \mu \bar{e}\right)_{\rm expt.} 
	\equiv \frac{\Gamma \left(K_L \to \mu^{\pm} e^{\mp}\right)}
                    {\Gamma \left(K_L \to {\rm all} \right)}  < 4.7 \times 10^{-12} 
\label{rareexpt}
\ee

\begin{figure}[h!]
\centering
\includegraphics[scale=0.75]{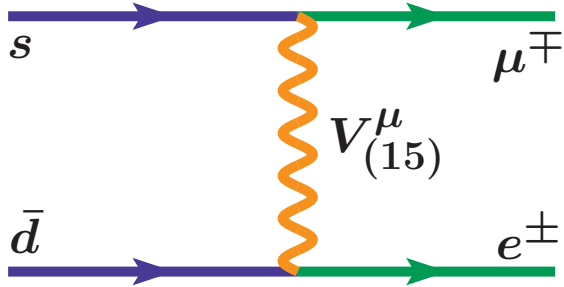}
\caption{Feynman diagram for rare kaon decays $K^0_L \to \mu^{\pm} e^{\mp}$ mediated by a heavy lepto-quark gauge boson of $SU(4)_C$ gauge symmetry.}
\label{fig:rarek}
\end{figure}
The leptoquark gauge bosons of $SU(4)_C$ in the adjoint representation $(1,1,15)$ 
under $G_{224}$ mediate rare kaon decay $K_L \to \mu^{\pm}  e^{\mp}$ whose 
Feynman diagram is shown in the Figure.\ref{fig:rarek}. Analytic formulas for the 
corresponding branching ratio is \cite{Deshpande:1982bf,Parida:1996xa},
\be
\mbox{Br.}\left(K_L \to \mu \bar{e}\right)= \frac{4\pi^2\alpha_s^2(M_C)\, 
m_K^4\, R}{G_F^2\, \sin^2\theta_C m_{\mu}^2{(m_s+m_d)}^2\,M_C^4},  \label{rarebth}
\ee
where the factor $R$ includes renormalization effects on the quark 
masses $m_d$ or $m_s$ from $\mu=M_C$ down to $\mu=\mu_0= 1$ GeV 
through the $G_{2113}$, the SM and the $SU(3)_C$ gauge symmetries. 

\begin{figure}
\begin{center}
\includegraphics[scale=.5, angle=-90]{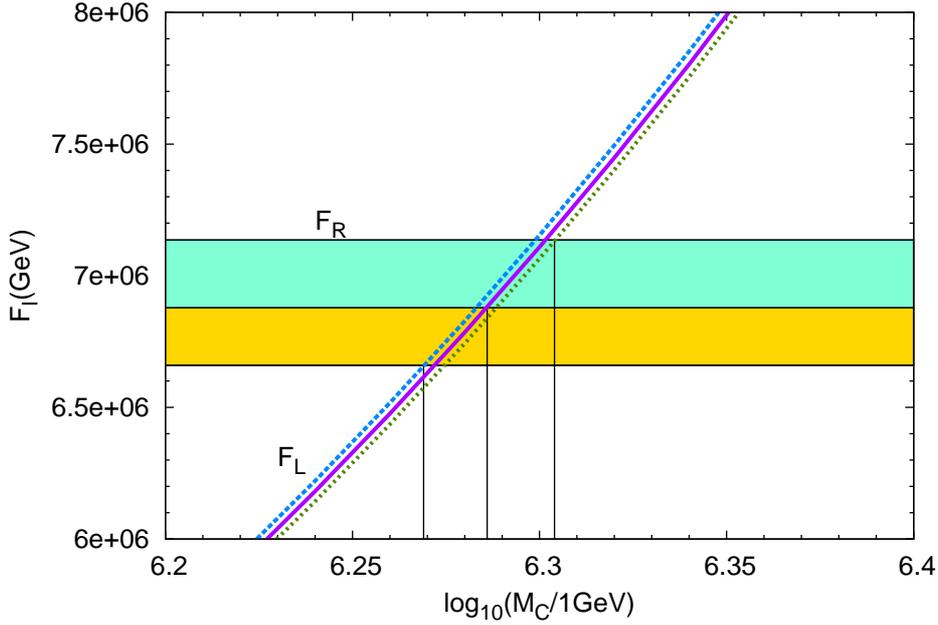}
\end{center}
\caption{Graphical representation of the method for numerical solution
of the lower bound on $M_C$. The horizontal lines are the RHS of the
inequality (\protect\ref{ineq2}) whereas the curve represents the LHS. The colored 
horizontal bands are due to uncertainties in the input parameters.}
\label{fig:mc_fl}
\end{figure}

Noting that the down quark or the strange quark mass satisfies 
the following renormalization group equations, 
\bea
m_{d,s}(M_C)&=&\frac{m_{d,s}(\mu_0)}{\eta_{\rm em}}R_{2113}
R_{213}^{(6)}R_{213}^{(5)}R_{\rm QCD}^{(5)}
R_{\rm QCD}^{(4)}R_{\rm QCD}^{(3)}
\eea
where
\begin{eqnarray}
& &R_{2113}=\Pi_{i} \left(\frac{\alpha_i(M_C)}
{\alpha_i(M_R^0)}\right)^{- C^i_1/{2 {a}^{(1)}_i}}\quad ,
\,i=2L,1R,B-L,3_C\, ,\nonumber\\
& &R^{(6)}_{213}=\Pi_{i}[\frac{\alpha_i(M_{R^0})}{\alpha_i(m_t)}]^{- C^i_2/{2{a^{(2)}_i}}},\quad 
R^{(5)}_{213}=\Pi_{i}[\frac{\alpha_i(m_t)}{\alpha_i(M_Z)}]^{-{C^i_2/{2a^{(3)}_i}}} \quad ,\,i=2L,Y,3C\, , 
\nonumber\\
& &{R^{(5)}}_{\rm QCD}=[\frac{\alpha_{S}(M_Z)}{\alpha_{S}(m_b)}]^{-{4/a^{(4)}}}\, , \quad 
{R^{(4)}}_{\rm QCD}=[\frac{\alpha_{S}(m_b)}{\alpha_i(m_c)}]^{- 4/a^{(5)}}
\nonumber\\
& &{R^{(3)}}_{QCD}=[\frac{\alpha_{S}(m_c)}{\alpha_{S}(\mu^0)}]^{- 4/a^{(6)}},
\label{Rfactors}
\end{eqnarray}
where the input parameters used in above eq.\,(\ref{Rfactors}) are: 
$C^i_1 = (0,0,1/4,8)$, $C^i_2 = (0,-1/5,8)$ and the one-loop 
beta-coefficients relevant for our present work are 
$a^{(1)}_i=(-3,{57/12}, 37/8, -7)$, $a^{(2)}_i=(-19/6, 41/10, -7)$, 
$a^{(3)}_i=(-23/6,{103/30}, -23/3)$, $a^{(4)}=-23/3$, $a^{(5)}=-25/3,~a^{(6)}=-9$. 
Now we can obtain the renormalization factor in eq.(\ref{rarebth}) 
\bea
 R&=&\left[R_{2113}R^{(6)}_{213}R^{(5)}_{213}R^{(5)}_{QCD}
 R^{(4)}_{QCD}R^{(3)}_{QCD}\right]^{-2}.\label{Rformula}
\eea
Using eq.(\ref{Rfactors}) and eq.(\ref{Rformula}) and 
eq.(\ref{rareexpt}), we derive the following inequality,
\bea
F_L(M_C,M_R^0)&>&\left[\frac{4\pi^2m_K^4R_p}{G_F^2\sin^2\theta_C
m_{\mu}^2{(m_s+m_d)}^2}\times 10^{11.318}\right]^{1/4},\label{ineq2}
\eea
where 
\bea
F_L(M_C,M_R^0)&=&M_C\alpha^{-3/14}_S(M_C)\alpha_Y^{-1/82}(M_{R^0})
\left[\frac{\alpha_{B-L}(M_C)}{\alpha_{B-L}(M_{R^0})}\right]^{-1/74}
\alpha_Y^{1/82}(m_t)\alpha_C^{-2/7}(m_t),\nonumber\\
R_p&=&\left[R_{213}^{(5)}R_{QCD}^{(5)}R_{QCD}^{(4)}R_{QCD}^{(3)}\right]^{-2}.\label{eq:FLRpdef}
\eea
In Fig.~{\ref{fig:mc_fl}} the function $F_L(M_C, M_R^0)$ in the LHS of 
eq.~(\ref{ineq2}) is plotted against $M_C$ for a fixed value of $M_R^0=5$~TeV, 
where the Horizontal lines represent the RHS of the same equation including
uncertainties in the parameters. 
Thus, for the purpose of this numerical estimation, keeping 
$M_R^0$ fixed at any value between $5-10$ TeV, we vary $M_C$ 
until the LHS of eq.(\ref{ineq2}) equals its RHS.
 
For our computation at $\mu_0=1$~GeV, we use the 
inputs $m_K=0.4976$~GeV, $m_d=4.8^{+0.7}_{-0.3}$~MeV, 
$m_s=95\pm 5$~MeV, $m_{\mu}=105.658$~MeV,  
$G_F=1.166\times 10^{-5}$~GeV$^{-2}$, and $\sin\theta_C=0.2254\pm 0.0007$, 
$m_b=4.18\pm 0.03$~GeV, $m_c=1.275\pm 0.025$~GeV, $m_{t}=172$ GeV. 
At $\mu=M_Z$ we have used $\sin^2\theta_W=0.23166 \pm 0.00012$, 
$\alpha_S=0.1184\pm 0.0007$, $\alpha^{-1}=127.9$ and utilized 
eq.(\ref{rareexpt})$-$eq.(\ref{eq:FLRpdef}). 
With $M_{R^0}=5$ TeV and $M_{Z'}\simeq 1.2$~TeV, the existing 
experimental upper bound on {\rm Br}.(${\rm K}_L\to \mu^{\mp} 
{e^{\pm}}$) gives the lower bound on the $G_{224}$ symmetry breaking 
scale 
\be
 M_C > (1.932^{+0.082}_{-0.074})\times 10^{6} {\rm GeV}.\label{eq:boundmc}
\ee
Noting from Fig.\ref{fig:unify} that in our model $\alpha_S(M_C)=0.0505$, we get 
from eq.(\ref{eq:boundmc}) as rare-kaon decay constraint on the $SU(4)_C$-leptoquark
gauge boson mass
\be
 M_{\rm lepto}  > (1.539^{+0.065}_{-0.059})\times 10^{6} {\rm GeV}.\label{eq:bound_lepto}
\ee
where the uncertainty is due to the the existing 
uncertainties in the input parameters. 
From the derived solutions to RGEs for gauge couplings this lower
bound on the lepto-quark gauge boson mass is easily accommodated in
our model.

\begin{figure*}[htb!]
\begin{minipage}[t]{0.48\textwidth}
\hspace{-0.4cm}
\begin{center}
\includegraphics[scale=.40, angle=0]{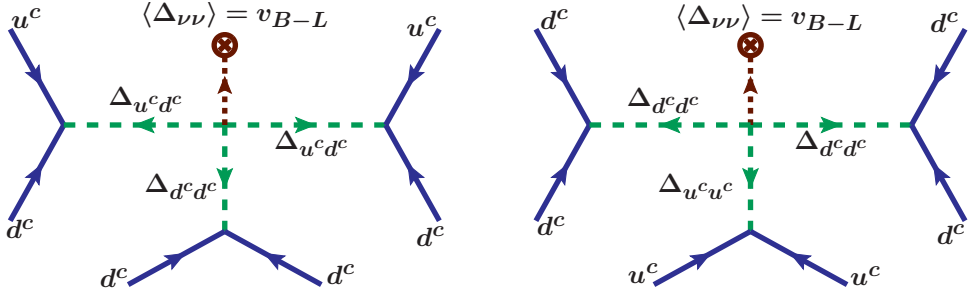}
\end{center}
\end{minipage}
\caption{Feynman diagrams for neutron-antineutron oscillation via 
	mediation of two $\Delta_{u^c d^c}$ and one $\Delta_{d^c d^c}$ 
         di-quark Higgs scalars as shown in the left-panel while mediation 
         of two $\Delta_{d^c d^c}$ and one $\Delta_{u^c u^c}$ 
         di-quark Higgs scalars as shown in the right-panel.} 
\label{fig:feyn-nnbar}
\end{figure*}

\subsection{\small Neutron-Antineutron Oscillation}
 Here we discuss the prospect of this model predictions for
 experimentally observable $n-{\bar n}$ oscillation while satisfying
   the rare-kaon decay constraint by fixing the $G_{224}$ symmetry
   breaking scale at $M_C\sim 2\times 10^{6}$~GeV as derived in eq.(\ref{eq:boundmc}). 
The Feynman diagrams for the  $n-\bar{n}$ oscillation processes are 
shown in left- and right-panel of Fig.\ref{fig:feyn-nnbar} 
where $\Delta_{\rm u^cu^c}$, $\Delta_{\rm d^cd^c}$, and 
$\Delta_{\rm u^cd^c}$ denote different diquark Higgs scalars 
contained in $\Delta_R (1, 3,\overline{10})\subset {126}_H$. 
The amplitude for these two diagrams can be written as
\be  
\mbox{Amp}^{(a)}_{n-{\bar n}}= \frac{f_{11}^3\lambda v_{B-L}}
{M_{{\Delta_{\rm u^cd^c}}}^4M_{{\Delta_{\rm d^cd^c}}}^2}, 
\quad \mbox{Amp}^{(b)}_{n-{\bar n}}= \frac{f_{11}^3\lambda v_{B-L}}
{M_{{\Delta_{\rm d^cd^c}}}^4\,M_{{\Delta_{\rm u^cu^c}}}^2},
\label{ampfig2a} \\
\ee
where $f_{11}=\left(f_{\Delta_{u^c d^c}}\right)
=\left(f_{\Delta_{d^c d^c}}\right)=\left(f_{\Delta_{u^c u^c}}\right)$ 
from the $SO(10)$ invariance and the quartic coupling 
between different di-quark Higgs scalar has its natural value i.e, 
$\mathcal{O}(0.1) - \mathcal{O}(1)$. 

The  $n-\bar{n}$ mixing mass element $\delta m_{\rm n{\bar n}}$ and 
the dibaryon number violating amplitude $W_{(B=2)}= \mbox{Amp}^{(a)}+\mbox{Amp}^{(b)}$ 
are related up to a factor depending upon combined effects 
of hadronic and nuclear matrix element effects
\be
\delta m_{\rm n{\bar n}}= \left( 10^{-4}\, \mbox{GeV}^6 \right) \cdot W_{B=2} .
\label{qcdcorramp}  
\ee
The experimentally measurable mixing time $\tau_{\rm n{\bar n}}$ is 
just the inverse of $\delta m_{\rm n{\bar n}}$
\be
\tau_{\rm n{\bar n}}=\frac{1}{\delta m_{\rm n{\bar n}}}.
\label{mixingtime}  
\ee
With $v_{B-L}=5$ TeV in the degenerate case, when all diquark 
Higgs scalars have identical masses $M_{\Delta}=10^5$ GeV, 
the choice of the parameters $f_{11}\simeq \lambda \sim \mathcal{O}(0.1)$ 
gives  $\tau_{\rm n{\bar n}}=6.58 \times 10^{9}$ 
sec. As described below our SO(10) model can fit all charged fermion 
masses and CKM mixings at the GUT scale  with two kinds 
of structures: (i) only one ${126}_H$, and (ii) two Higgs representations  
${126}_H$ and ${126}_H^{\prime}$. In the minimal  case the Yukawa coupling 
$f$ of ${126}_H$ to fermions has a diagonal structure,
\be
f={\rm diag}(0.0236, -0.38, 1.5), \label{yukcasei}
\ee
which gives through eq.\,(\ref{ampfig2a}),  eq.\,(\ref{qcdcorramp}), 
and eq.\,(\ref{mixingtime})
\be  
\tau_{n-{\bar n}} = 10^8-10^{10} {\rm  secs}.\label{casei}
\ee
This model prediction  is accessible to ongoing search experiments
\cite{nnbar-osciln-expt}. However, the GUT scale fit to the fermion 
masses can be successfully implemented without constraining the $f$ 
values when a second ${126}_H^{\prime}$ is present at the GUT scale 
with all its component at $M_U$ except $\xi^{\prime}(2,2,15)$ being
around the $M_P$ scale. Then using 
$f_{11}=0.1-0.01$, the estimated value turns out to be
\be 
\tau_{n-{\bar n}} \sim 10^9-10^{13} {\rm  sec}.\label{caseii}
\ee
Out of this the mixing time in the renge $10^9-10^{10}$ sec can be probed by
ongoing experiment \cite{nnbar-osciln-expt}.
\begin{table}
\begin{center}
\begin{tabular}{|l|l|c|c|l|}
\hline
$f_{11}$& $\lambda$& $M_{\Delta_{u^cd^c}}$~(GeV)& $M_{\Delta_{d^cd^c}}$~(GeV) 
& $\tau_{{\rm n}-\bar{\rm n}}$~(secs) \\
\hline
0.1 & 0.1 & $10^5$ & $10^5$ & $6.6\times 10^9$ \\
\hline
0.0236 & 0.1 & $10^5$ & $10^5$ & $2.5\times 10^{13}$ \\
\hline
0.0236 & 1.0 & $10^5$ & $10^5$ & $2.5\times 10^{14}$ \\
\hline
0.1 & 0.1 & $10^4$ & $10^5$ & $6.6\times 10^9$ \\
\hline
0.0236 & 1.0 & $10^4$ & $10^5$ & $2.5\times 10^{13}$ \\
\hline
0.0236 & 1.0 & $10^5$ & $10^4$ & $2.5\times 10^{13}$ \\
\hline
\end{tabular}
\caption{Predictions for n$-\bar{\rm n}$ oscillation mixing time as a function 
of allowed couplings and masses of diquark Higgs scalars in the model described
in the text.}\label{tab:n_n_mixing}
\end{center}
\end{table}

\section{{\small DETERMINATION OF DIRAC NEUTRINO MASS MATRIX}}\label{sec:4}
The Dirac neutrino mass near the TeV scale forms an essential 
ingredient in the estimations of inverse seesaw contribution 
to light neutrino masses and mixings as well as the LFV and LNV 
processes in this model in addition to predicting leptonic 
CP-violation through non-unitarity effects. Since the procedure 
for determination of $M_D$ has been discussed earlier 
\cite{Awasthi:2013ff}, we mention it briefly here in the 
context of the present model. In order to 
obtain the  Dirac neutrino mass matrix $M_D$ and the RH Majorana 
mass matrix $M_N$ near TeV scale, at first the PDG values 
\cite{Yao:2006} of fermion masses at the electroweak scale are 
extrapolated to the GUT scale using the renormalization group equations 
(RGEs) for fermion masses in the presence of the SM for $\mu=M_Z - 5$ TeV, 
and from $\mu=5-10$ TeV using the RGEs in the presence of $G_{2113}$ 
symmetry \cite{dp:2001,ap:2011aa}. From $\mu=5-100$ TeV, RGEs corresponding to two 
Higgs doublets in the presence of $G_{2113}$ symmetry are used 
\cite{ap:2011aa}. These two doublets which act like up-type 
and down type doublets are treated to have originated from separate 
representations ${10}_{H_1}$ and ${10}_{H_2}$  of $SO(10)$. For mass 
scale $\mu \ge 10^5$ GeV till the GUT scale the fermion mass RGEs
derived in the presence of the $G_{224}$ and $G_{224D}$ symmetries 
\cite{fuku:2002} are exploited. Then at the GUT scale $\mu=M_{\rm GUT}$ 
we obtain the following values of mass eigenvalues and the CKM mixing matrix
$m_u^0=1.301$~ MeV, $m_c^0=0.1686$~GeV, $m_t^0=51.504$~GeV, 
$m_d^0=1.163$~MeV, $m_s^0=23.352$~ MeV, $m_b^0=1.0256$~GeV, 
$m_e^0=0.2168$~MeV, $m_{\mu}^0=38.846$~MeV, 
~~$m_{\tau}^0=0.962$ GeV,\\
\begin{eqnarray}
{V^0_{\rm CKM}(M_{\rm GUT})}&=& 
{ \begin{pmatrix} 0.976& 0.216& -0.0017-0.0035i\\
-0.216-0.0001i& 0.976-0.0000i& 0.0310\\
0.0083-0.0035i& -0.03-0.0007i& 0.999\end{pmatrix}},
\label{eqn:vckmu} 
\end{eqnarray}

Formulas for different fermion mass matrices at the GUT scale 
have been discussed in \cite{ap:2011aa,Awasthi:2013ff}
\begin{eqnarray}
M_u^0 &=& G_u+F,~ M_D^0 = G_u-3F,\nonumber\\
M_d^0 &=& G_d+F, ~ M_l=G_d-3F, \label{fformulas}
\end{eqnarray}
where $G_u=Y_1v_u, G_d=Y_2v_d$ , and in the absence of ${126}_H$ 
in those models , the diagonal structure of $F$ was shown to 
originate from available non-renormalizable higher dimensional
operators.The new interesting point here is that the present 
model permits $F$ to be renormalizable using the ansatz 
 \cite{Babu:1992ia} $F=fv_{\xi}$, and the induced VEV $v_{\xi}$ 
 of $\xi(2,2,15)\subset {126}_H$ is predicted within the 
 allowed mass scales of the $SO(10)$ while safeguarding 
 precision gauge coupling unification.\\
 
Using the charged-lepton diagonal mass basis and eq.(\ref{fformulas}) we have
\bea
M_e(M_{GUT})&=&{\rm diag}(0.000216,0.0388,0.9620)~{\rm GeV},\nonumber\\
 G_{d,ij} &=& 3F_{ij},~(i \neq j).\label{hdij}
\eea
In the present model, type-II seesaw contribution being negligible
 and the neutrino oscillation data being adequately represented by 
 inverse seesaw formula, there is no compelling reason 
 for the Majorana coupling $f$ to be non-diagonal. On the other 
 hand diagonal texture of RH neutrino mass matrix has been 
 widely used in the literature in a large class of $SO(10)$ 
 models. Moreover, as we see below, the diagonal structure 
of  $f$ which emerges in the minimal model 
 exactly predicts the RH neutrino masses accessible to LHC and 
 the neutrino oscillation data.\footnote{Alternatively the 
 fermion masses at the GUT scale can be fitted by the diagonal 
 coupling $f^{\prime}$ of a second ${126}_{H^{\prime}}$ whose 
 $\xi^{\prime}(2,2,15)$ component can be fine-tuned to have mass at the 
 same intermediate scale to provide the desired VEV. In this 
 case the $f$ and RH Majorana neutrino mass matrix $M_N$ is 
 allowed to possess a general $3\times 3$ matrix structure withou any
 apriori constraint.}
 We then find that diagonal texture of $f$ gives 
the matrix $G_d$ to be also diagonal leading to the relations
\bea
\mbox{G}_{\rm d,ii} + \mbox{F}_{\rm ii} &=& m^0_i,~(\mbox{i=d,s,b}),\nonumber\\
\mbox{G}_{\rm d,jj} -3 \mbox{F}_{\rm jj}&=& m^0_j,~(j=e,\mu,\tau).\label{hfrel}
\eea
\bea
F&=&{\rm diag}{1\over 4}(m^0_d-m^0_e, m^0_s-m^0_{\mu}, m^0_b-m^0_{\tau}),\nonumber\\
&=&{\rm diag}(2.365\times 10^{-4}, -0.0038, +0.015)~{\rm GeV},\nonumber\\
G_d&=&{\rm diag}{1\over 4}(3m^0_d+m^0_e, 3m^0_s+m^0_{\mu}, 3m^0_b+m^0_{\tau}),\nonumber\\
&=&{\rm diag}(9.2645\times 10^{-4}, 0.027224, 1.00975)~{\rm GeV},\label{FGd}
\eea
where we have used the RG extrapolated values at the GUT scale. 
It is clear from the value of the mass matrix $F$ in eq.\,(\ref{FGd}) 
that we need a small VEV  $v_{\xi}\sim 10$ MeV to fit the fermion mass 
fits at the GUT scale. To verify that this $v_{\xi}$ is naturally obtained 
in this model, we note that the spontaneous symmetry breaking  in this
model $G_{224}\rightarrow G_{2113}$ occurs through the VEV of 
$(1,3,15)_H\subset 210_H$. Then the desired trilinear term in the scalar potential $V$ 
gives the natural value of the VEV  
\bea
 V &=& \lambda_3 M_{GUT} {210}_H.{126}_H^{\dagger}{10}_H \nonumber\\  
   &&=\lambda_3M_{GUT}(1,3,15)_{210}.(2,2,15)_{126}.(2,2,1)_{{10}_{1,2}},\nonumber\\
v_{\xi}&\sim& \lambda_3M_{GUT}M_Cv_{\rm ew}/M_{\xi}^2 
       = 10 {\rm MeV} - 100 {\rm MeV}, \label{inducedvev}
\eea
for $M_{\xi}= 10^{12}-10^{13}$ GeV. 

Repeating the RG analysis of ref.\cite{Awasthi:2013ff} we have verified that the precision gauge coupling 
unification is unaffected when $\xi(2,2,15)$ occurs at such high 
intermediate scales except for an increase of the GUT scale by nearly 
$2$ and the GUT fine structure constant by nearly three times. That 
the Parity violating scale and the GUT scale would be marginally 
affected is easy to understand because the contribution due to $\xi(2,2,15)$ 
to all the three one-loop beta-function coefficients are almost similar 
$\delta b_{\rm 2L} = \delta b_{\rm 2R} =5, \delta b_{\rm 4C}=5.333$. 
That the unification is bound to occur can be easily seen because there 
are only two gauge coupling constant lines for $\mu > M_P$. 
   
Using the computed values of $M_u^0$  and the value of $F$ from
eq.\,(\ref{FGd})in eq.\,(\ref{fformulas}), 
 gives the the  matrix $G_u$ at $\mu=M_{GUT}$. 
Another by product of this fermion mass fit at the GUT scale is that 
the matrix elements of $F$ now gives $f={\rm diag}(f_1,f_2,f_3)$ and 
consequently the RH neutrino mass hierarchy 
$M_{N_1}:M_{N_2}:M_{N_3}=0.023:-0.38:1.5$. This hierarchy is consistent 
with lepton-number and lepton flavor violations discussed in Sec.2, 
Sec.4, and Sec.5    
\bea
{\small G_u(M_{GUT})}&=&{\small \begin{pmatrix}
0.0095         & 0.0379-0.0069i  & 0.0635-0.1671i\\
0.0379+0.0069i & 0.2637           & 2.117+0.0001i\\
0.0635+0.1672i  & 2.117-0.0001i  & 51.444 \end{pmatrix}}\,
{\small {\rm GeV}} .\label{GuU}
\eea

Now using eq.\,(\ref{FGd}) and eq.\,(\ref{GuU}) in eq.\,(\ref{fformulas}) 
gives the Dirac neutrino mass matrix $M_D$ at the GUT scale
\bea
{\small M^0_D(M_{GUT})}&=&{\small \begin{pmatrix}
0.00876          & 0.0380-0.0069i   & 0.0635-0.1672i \\
0.0380+0.0069i  & 0.3102            & 2.118+0.0001i\\
0.0635+0.1672i   & 2.118-0.0001i   & 51.63
\end{pmatrix}}~{\small {\rm GeV}}.\label{mdU}
\eea 

Noting that $F=fv_{\xi}={\rm diag}(f_1,f_2,f_3)v_{\xi}$ in eq.(\ref{FGd}), $v_{\xi}=10$\,MeV 
gives $(f_1,f_2,f_3)=(0.0236, -0.38, 1.5)$. Then the allowed 
solution to RGEs for gauge coupling unification with $M_R^0=v_R=5$ TeV 
gives $M_{N_1}=115$ GeV, $M_{N_2}=1.785$ TeV, and $M_{N_3}=7.5$ TeV. While 
the first RH neutrino is lighter than the current experimental limit 
on $Z_R$ boson mass, the second one is in-between the $Z_R$ and $W_R$ 
boson mass limits , but the heaviest one is larger than the $W_R$ mass 
limit. These are expected to provide interesting collider signatures 
at LHC and future accelerators. This hierarchy of the RH neutrino masses 
has been found to be consistent with lepton-number and lepton flavor 
violations discussed in Sec.2, Sec.4, and Sec.5. Then following the 
top-down approach we obtain the value of $M_D$ at the TeV scale 
\begin{eqnarray}
\label{eq:md-mr0}
M_D(M_R^0) =  \left( \begin{array}{ccc} 
 0.02274          & 0.0989-0.0160i    & 0.1462-0.3859i\\
 0.0989+0.0160i & 0.6319              & 4.884+0.0003i\\
 0.1462+0.3859i   & 4.884-0.0003i    & 117.8
\end{array} \right) \text{GeV}\,.
\end{eqnarray}

We will use $M_N=(0.115, 1.7, 7.8)$~TeV and the $M_D$ matrix of
eq.(\ref{eq:md-mr0}) to predict LFV and LNV decays in the next two
sections. 


\section{{\small FITTING THE NEUTRINO OSCILLATION DATA BY 
GAUGED INVERSE SEESAW FORMULA}}\label{sec5}
In the presence of three singlet fermions $S_i,\,(i=1,2,3)$, 
the inverse seesaw mechanism \cite{PhysRevLett.56.561, 
PhysRevD.34.1642, Wyler:1982dd, ap:2011aa,Awasthi:2013ff} 
is implemented in the present model through the $SO(10)$ invariant 
Yukawa Lagrangian that gives rise to the $G_{2113}$ invariant interaction 
near the TeV scale \cite{ap:2011aa,Awasthi:2013ff} 
where $\chi_R(1,1/2,-1,1) \subset {16}_H$ generates the $N-S$ mixing term,
\begin{eqnarray}
\mathcal{L}_{\rm Yuk}&= & Y^a{\bf {16}}.{\bf {16}}.{10}^a_H 
+ f{\bf {16}}.{\bf {16}}.{126^\dagger}_H + y_{\chi} {\bf {16}}.{\bf {1}}.{16^\dagger}_H 
+{\bf {\mu_S}}{\bf {1}}.{\bf {1}}      \nonumber\\
&& \supset Y^{\ell} \overline{\ell}_L\, N_R\, \Phi_1 + f\, N^c_R\, 
N_R \Delta_R + F\, \overline{N}_R\, S\, \chi_R 
+S^T \mu_S S +\text{h.c.}.\nonumber 
\end{eqnarray}
This Lagrangian  gives rise to the $9\times 9$ neutral fermion 
mass matrix after electroweak symmetry breaking.\\
\begin{equation}
\mathcal{M}= \left( \begin{array}{ccc}
                0        & 0 & M_D   \\
              0 &    \mu_S         & M \\
              M^T_D & M^T & M_N
                      \end{array} \right) \, ,
\label{eqn:numatrix}       
\end{equation}
In contrast to the SM where all three matrices $M_N, M$, and $\mu_S$ 
have no dynamical origins, in this model the first two have 
dynamical interpretations $M_N=fv_R$, $M=y_{\chi}v_{\chi}$; 
only $\mu_S$ suffers from this difficulty.
   
In this model the RH neutrinos being heavier than the other two 
fermion mass scales in the theory with $M_N \gg M > M_D, \mu_S$, 
they are at first integrated out from the Lagrangian, which, in 
the $\left(\nu,~ S\right)$ basis, gives the $6 \times 6$ mass matrix
\begin{eqnarray}
\mathcal{M}_{\rm eff} =- \left( \begin{array}{cc}
    M_DM_N^{-1} M^T_D  &  M_D M_N^{-1} M^T \\
    MM_N^{-1} M_D^T       & MM_N^{-1} M^T - \mu_S 
        \end{array} \right) \, ,
\label{eqn:eff_numatrix}       
\end{eqnarray}
This is further block diagonalised to find that the would be 
dominant $type-I$ seesaw contribution completely cancels out leading 
to the gauged inverse mass formula for light neutrino mass matrix 
and also another formula for the sterile neutrinos(S) 
\bea 
m_\nu&=&M_DM^{-1}\mu_S (M_DM^{-1})^T \label{numas:inverse}\\
m_{S}&=&\mu_S-MM_N^{-1}M^T \label{numas:sterile}
\eea
The complete $6 \times 6$ unitary mixing matrix which diagonalizes the 
light-sterile neutrino effective mass matrix $\mathcal{M}_{\rm eff}$ is
\bea 
\mathcal{V}_{6 \times 6}& =& \mathcal{W} \cdot \mathcal{U} \nonumber \\
&=&
\bmt 
1-\frac{1}{2}XX^\dagger & X \\
-X^\dagger & 1-\frac{1}{2}X^\dagger X 
\emt \cdot \,
\bmt
U_{\nu} & 0 \\
0       & U_{S}
\emt 
\eea
In this extended inverse seesaw scheme, the light neutrinos are actually 
diagonalized by a matrix which is a part of the full $6\times 6$ 
mixing matrix $\mathcal{V}_{6 \times 6}$ 
\begin{eqnarray}
\mathcal{N} &\simeq& \left(1-\frac{1}{2} X\, X^\dagger \right) 
U_{\rm PMNS} =  \left(1- \eta \right) U_{\rm PMNS}
\end{eqnarray}
where $\eta = \frac{1}{2} M_D M^{-1}\, (M_D M^{-1})^\dagger $ is a 
measure of non-unitarity contributions. In the $(\nu,S,N)$ basis, 
adding RH Majorana mass $M_N$ to eq.(\ref{eqn:eff_numatrix}), the complete 
mixing matrix \cite{Grimus:2000vj,Awasthi:2013ff} 
diagonalizing the resulting  $9 \times 9$ neutrino mass matrix  turns out to be
\begin{eqnarray}
\mathcal{V}&\equiv&
\bmt 
{\cal V}^{\nu\hat{\nu}}_{\alpha i} & {\cal V}^{\nu{\hat{S}}}_{\alpha j}
 & {\cal V}^{\nu \hat{N}}_{\alpha k} \\
{\cal V}^{S\hat{\nu}}_{\beta i} & {\cal V}^{S\hat{S}}_{\beta j} 
& {\cal V}^{S\hat{N}}_{\beta k} \\
{\cal V}^{N\hat{\nu}}_{\gamma i} & {\cal V}^{N\hat{S}}_{\gamma j} 
& {\cal V}^{N\hat{N}}_{\gamma k} 
\emt \nonumber \\
&=&\bmt 
\left(1-\frac{1}{2}XX^\dagger \right) U_\nu  & 
\left(X-\frac{1}{2}ZY^\dagger \right) U_{S} & 
Z\,U_{N}     \\
-X^\dagger\, U_\nu   &
\left(1-\frac{1}{2} \{X^\dagger X + YY^\dagger \}\right) U_{S} &
\left(Y-\frac{1}{2} X^\dagger Z\right) U_{N}   \\
y^*\, X^\dagger\, U_{\nu} &
-Y^\dagger\, U_{S} &
\left(1-\frac{1}{2}Y^\dagger Y\right)\, U_{N} 
\emt \, ,
 \label{eqn:Vmix-extended}
\end{eqnarray}
as shown in the appendix. In eqn.\,(\ref{eqn:Vmix-extended}) 
$X = M_D\,M^{-1}$, $Y=M\, M^{-1}_N$, $Z=M_D\,M^{-1}_N$, and $y=M^{-1}\,\mu_S$.

Although the $N-S$ mixing matrix $M$ in general can be non diagonal, 
we have assumed it to be diagonal partly to reduce the unknown 
parameters and as we shall see the LFV effects constrain the 
diagonal elements. Noting that $\eta_{\alpha \beta} = \frac{1}{2} 
\sum^{3}_{k=1}\, (M_{D_{\alpha k}}\,M^*_{D_{\beta k}})/{M^2_{k}}$, 
the entries of the $\eta$ matrix are constrained from various 
experimental inputs like e.g. rare leptonic decays, invisible 
Z-boson width, neutrino oscillations etc. For illustration 
let us quote the bound on these elements of $\eta$ on $90 \%$ C.L. 
\footnote{For related references on the $90 \%$ C.L of the bounds on
  the elements $|\eta_{\alpha\beta}|$ see references cited in \cite{ap:2011aa,Awasthi:2013ff}.} 
$|\eta_{ee}| \leq 2.0 \times 10^{-3}$, $|\eta_{\mu \mu}| \leq 8.0\times 10^{-4}$, 
$|\eta_{\tau \tau}| \leq 2.7 \times 10^{-3} $, $|\eta_{e\mu}| \leq 3.5 \times 10^{-5}$, 
$|\eta_{e \tau}| \leq 8.0 \times 10^{-3}$, and $|\eta_{\mu \tau}| \leq 5.1 \times 10^{-3}$. 
Whereas the possible CP phases of the elements of $\eta_{\alpha \beta}$ 
($=\phi_{\alpha \beta}$) are not constrained, the knowledge of 
$M_D$ matrix given in eq.(\ref{eq:md-mr0}) and saturation 
of the lower bound on $|\eta_{\tau \tau}| = 2.7 \times 10^{-3}$ 
leads to a relation between diagonal elements of M, 
\begin{eqnarray}
\frac{1}{2}\, \bigg[\frac{0.17}{M^2_1}+\frac{23.853}{M^2_2}
+\frac{13876.84}{M^2_3} \bigg] = 2.7 \times 10^{-3}
\label{eqn:rel-eta}
\end{eqnarray}
The above relation can be satisfied by the partial degenerate values 
of $M$ as $M_1 = M_2 \geq 100$ GeV and $M_3 \geq 2.15$ TeV while it also accommodates 
the complete non-degenerate values $M_1 \geq 10$ GeV, $M_2 \geq 120$ GeV, and $M_3 
\geq 2.6$ TeV. For degenerate $M$, this gives  $M_1 = M_2 = M_3 = 1.6$ TeV. 
The elements of $\eta$ can be different for different values of $M$ allowed in 
our model. We need to know the PMNS mixing matrix  and $\eta$ in order to 
estimate the non-unitarity leptonic mixing matrix  $\mathcal{N}_{3 \times 3}$. 

Our analysis carried out for a normal hierarchy (NH) of light neutrino 
masses can be repeated also for inverted hierarchical (IH) or for 
quasi-degenerate (QD) masses to give correspondingly different values 
of the ${\mu}_S$ matrix. For example, using NH for which 
$\hat{m}_{\nu}^{\rm diag} = {\rm diag}(\text{0.00127 }, ~\text{0.00885 }, 
~\text{0.0495 })$~eV consistent with the central values of a recent 
global analysis of the neutrino oscillation parameters \cite{Fogli}
$\Delta m^2_{\rm sol}=  7.62\times 10^{-5}~{\rm eV}^2,~
\Delta m^2_{\rm atm} = 2.55\times 10^{-3}~{\rm eV}^2,~
\theta_{12}=34.4^\circ,~ \theta_{23}=40.8^\circ,\theta_{13}=9.0^\circ,
\delta=0.8\pi\;$
and assuming vanishing Majorana phases $\alpha_1=\alpha_2=0$ , 
we use the non-unitarity mixing matrix $\mathcal{N} = 
\left(1- \eta \right) U_{\rm PMNS}$, and  the relation 
$m_\nu = \mathcal{N} \hat{m}_\nu \mathcal{N}^T$, 
to derive the form of $\mu_S$ matrix from the light neutrino mass 
formula (\ref{numas:inverse}) 
\begin{eqnarray}
\mu_S &=& X^{-1}\, \mathcal{N} \hat{m}_\nu \mathcal{N}^T\, (X^T)^{-1} \nonumber \\
&=&\left(
\begin{array}{ccc}
 0.001+0.0004\,i & -0.0026-0.0012\,i & 0.0013 \\
 -0.0026-0.0012\,i & 0.0067+0.0023\,i & -0.0034 \\
 0.0013 & -0.0034 & 0.0014-0.0006\, i
\end{array}
\right)\mbox{GeV} 
\end{eqnarray}

\section{{\small LEPTON FLAVOR VIOLATIONS}}\label{sec6}
Within the framework of this extended seesaw scheme 
\cite{Awasthi:2013ff}, the dominant contributions 
are mainly through the exchange of heavy sterile neutrinos ($S$)
as well as heavy RH neutrinos ($N_R$) with branching ratio
\cite{Ilakovac:1994kj,Cirigliano:2004mv,ap:2011aa,Awasthi:2013ff,
Leontaris:1985qc,Ilakovac:2012sh}
\begin{eqnarray}
& &\text{Br}\left(\ell_\alpha \rightarrow \ell_\beta + \gamma \right) =
          \frac{\alpha^3_{\rm w}\, s^2_{\rm w}\, m^5_{\ell_\alpha}}
          {256\,\pi^2\, M^4_{W}\, \Gamma_\alpha} 
           \left|\mathcal{G}^{N}_{\alpha \beta} + \mathcal{G}^{S}_{\alpha \beta}\right|^2 
\label{eq:LFV} \\
&\text{where}~&\mathcal{G}^{N}_{\alpha \beta} =
        \sum_{k} \left(\mathcal{V}^{\nu\, \hat{N}}\right)_{\alpha\, k}\, 
         \left(\mathcal{V}^{\nu\, \hat{N}}\right)^*_{\beta\, k} 
         \mathcal{F}\left(\frac{m^2_{N_k}}{M^2_{W_L}}\right) 
         \nonumber \\
& & \mathcal{G}^{S}_{\alpha \beta} = \sum_{j} \left(\mathcal{V}^{\nu \hat{S}}\right)_{\alpha\, j}\, 
         \left(\mathcal{V}^{\nu \hat{S}}\right)^*_{\beta\, j} 
         \mathcal{F}\left(\frac{m^2_{S_j}}{M^2_{W_L}}\right)  \nonumber \\
&\text{with}& \mathcal{F}(x) = -\frac{2 x^3+ 5 x^2-x}{4 (1-x)^3} 
                - \frac{3 x^3 \text{ln}x}{2 (1-x)^4}\, .\nonumber
\end{eqnarray}
where the summation over $j$ and $k$ goes over number of sterile 
neutrinos $S_{j}$ and for heavy right-handed Majorana 
neutrinos $N_{k}$ and the mixing matrices are 
$\mathcal{V}^{\nu\hat{S}}_{\alpha\, j} = \{X\, U_S\}_{\alpha\, j}$ and 
$\mathcal{V}^{\nu\hat{N}}_{\alpha\, k} = \{Z\, U_N \}_{\alpha\, k}$ 
with $X=\frac{M_D}{M}$ and $Z=\frac{M_D}{M_N}$. 
The allowed ranges of $M_i,(i=1,2,3)$ from the LFV constraint 
eq.(\ref{eqn:rel-eta}) and the predicted values of $M_{N_i}, (i=1,2,3)$ 
now determine the mass eigen values of the sterile neutrinos 
leading to $M_{S_i}=\{12.5, 49, 345.6\}\, \mbox{GeV}$ for 
$M=\text{diag}[40, 300, 1661]$ GeV, and $M_N = \text{diag}[115, 1785,
  7500]$ GeV.

The neutrino mixing matrices are estimated numerically 
\begin{eqnarray}
&&\mathcal{N} \equiv \mathcal{V}^{\nu \hat{\nu}} =
\left(
\begin{array}{ccc}
 0.8143-0.0008i & 0.5588+0.0002i & 0.1270+0.0924i \\
 -0.3587-0.0501i & 0.6699-0.0343i & -0.6472-0.0001i \\
 0.4489-0.0571i & -0.4849-0.0394 & -0.7438-0.0001i
\end{array}
\right)\, ,\\
&& 
\mathcal{V}^{\nu \hat{S}}=\left(
\begin{array}{ccc}
 0.0542 & 0.0325-0.0052i & 0.0086-0.0227i \\
 0.2358+0.0380i & 0.2075 & 0.2869 \\
 0.3465+0.9159i & 1.597 & 6.920
\end{array}
\right)\times 10^{-2}i\,,\\
&& 
\mathcal{V}^{\nu \hat{N}}=\left(
\begin{array}{ccc}
 0.0170 & 0.0053-0.0009i & 0.0018-0.0048i \\
 0.0740+0.0119i & 0.0340 & 0.0608 \\
 0.1089+0.2865i & 0.2625 & 1.467
\end{array}
\right)\times 10^{-2}\, .
\end{eqnarray}
Compared to RH neutrinos, the branching ratios due to exchanges of 
sterile neutrino ($S_i$) are found to be more dominant 
\begin{eqnarray}
\text{Br}\left(\mu \rightarrow e + \gamma \right) &=& 
 3.5 \times 10^{-16}.\nonumber \\
\end{eqnarray}
 Similarly, other LFV decay amplitudes 
are  estimated leading to~\cite{Adam:2013mnn}
\begin{eqnarray}
& &\text{Br}\left(\tau \rightarrow e + \gamma \right) = 3.0 \times 10^{-14} \, ,
\nonumber \\
& &\text{Br}\left(\tau \rightarrow \mu + \gamma \right) = 4.1 \times
10^{-12} \, .
\end{eqnarray}
These branching ratios are accessible to ongoing search experiments

We have also noted here that the leptonic CP-violating parameter due
to non-unitarity effects is $J\simeq 10^{-5}$ which is similar to the
model prediction of ref.\cite{Awasthi:2013ff}. 
\section{{\small NEW CONTRIBUTIONS TO NEUTRINO-LESS  
DOUBLE BETA DECAY IN THE $W_L-W_L$ CHANNEL}}\label{sec7}
In the generic inverse seesaw, there is only one small lepton 
number violating scale $\mu_S$ and the lepton number 
is conserved in $\mu_S=0$ limit leading to vanishing nonstandard 
contribution to the $0\nu 2\beta$ transition amplitude. 
On the contrary, in the extended seesaw under consideration, 
it has been shown for the first time that there can be a new 
dominant contributions from the exchanges of  heavy sterile  
neutrinos \cite{Awasthi:2013ff}. The main thrust of our 
discussion will be new contribution arising from exchange 
of heavy sterile neutrinos $S_i$ with Majorana mass $M_S =\mu_S-
M(1/M_N)M^T$ as explained in Sec.\,\ref{sec5}.
Because of heavy mass of $W_R$ boson in this theory, the RH 
current contributions are damped out. 
\begin{figure}[htb]
\centering
\includegraphics[scale=0.5]{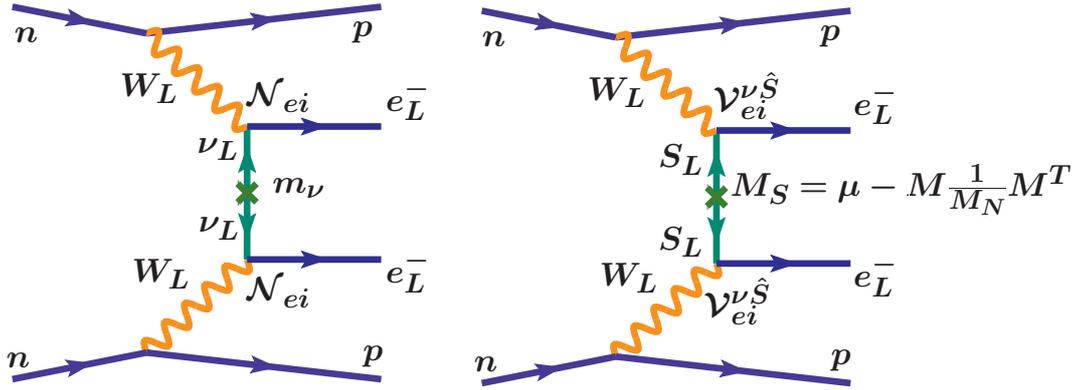}
\caption{$W^-_L - W^-_L$ mediated channel with light $\nu_i$ 
          and sterile $S_i$ Majorana neutrino exchanges.}
\label{fig:WLWL}
\end{figure}

In the mass basis, we have $\nu_\alpha = \mathcal{N}_{\alpha\, i}\, 
\nu_{m_i} + \mathcal{V}^{\nu\hat{S}}_{\alpha\, j}\, S_{m_j}$. 
In addition to the well known standard contribution in the  
$W^-_L - W^-_L$ channel shown in the left-panel of Fig.\,\ref{fig:WLWL}, 
we note the new contribution shown in the right-panel of Fig.\,\ref{fig:WLWL} 
with the corresponding amplitudes 
\begin{eqnarray}
\label{eq:amp_ll} 
& &\mathcal{A}^{LL}_{\nu} \propto G^2_F \frac{\left(\mathcal{V}^{\nu \hat{\nu}}_{e\,i}\right)^2\, m_{\nu_i}}{p^2} \,,\\
& &\mathcal{A}^{LL}_{S} \propto G^2_F \frac{\left(\mathcal{V}^{\nu\hat{S}}_{e\,j}\right)^2}{M_{S_j}} \, .
\end{eqnarray}
where $|p^2| \simeq \left(190~\mbox{MeV}\right)^2$ represents 
neutrino virtuality momentum and $G_F = 1.2 \times 10^{-5}\, 
\mbox{GeV}^{-2}$. 

\begin{figure}[htb!]
\begin{center}
\includegraphics[scale=1.0,angle=0]{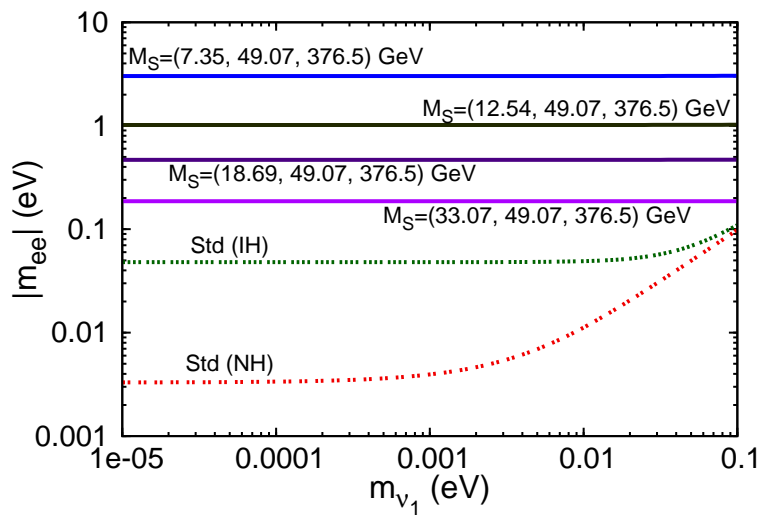}
\end{center}
\caption{Predicted effective mass parameter due to $W^-_L - W^-_L$ 
mediated channel with light $\nu_i$ and sterile $S_i$ Majorana 
neutrino exchanges. We have used best-fit oscillation parameters 
          while predicting standard contribution due to NH and 
          IH pattern of the light neutrino masses.}
\label{fig:eff-mass}
\end{figure}
Noting from eq.(\ref{eqn:Vmix-extended}) that 
$(\mathcal{V}^{\nu \hat{S}}_{e\,j})^2=(M_D/M)_{e\,j}^2$, the RHS of 
eq.(34) is expected to dominate because of three reasons:(i) 
Dirac neutrino mass origin from quark-lepton symmetry in $SO(10)$,
(ii) Smaller values of diagonal elements of the $N-S$ mixing matrix 
$M$, (iii) smaller eigen values of the heavy sterile Majorana neutrino mass: 
$M_S =\mu_S-M(1/M_N)M^T$. The mixing matrix elerments necessary for prediction 
of $0\nu \beta \beta$ amplitude can be represented as,
\begin{eqnarray}
& &\mathcal{N}_{e\, 1} = 0.819, \quad \mathcal{N}_{e\, 2} 
= 0.552, \quad \mathcal{N}_{e\, 3} = 0.156 \, ,\nonumber \\
& &\mathcal{V}^{\nu \hat{S}}_{e\, 1} = 0.00015, \quad \mathcal{V}^{\nu\hat{S}}_{e\, 2} 
= 0.00068, \quad \mathcal{V}^{\nu \hat{S}}_{e\, 3} = 0.00022 \, .
\end{eqnarray}

\subsection{New formula for half-life and bound on sterile neutrino mass}
We derive a new formula for half-life of $0\nu \beta \beta$ decay
as a function of heavy sterile neutrino masses and other parameters 
in the theory. We then show how the current experimental bounds 
limit the lightest sterile neutrino mass to be $M_{S_1}\ge 14 \pm 4$ GeV.

Using results discussed in previous sections, the inverse 
half-life is presented in terms of $\eta-$ parameters 
and others including the nuclear matrix elements 
\cite{Awasthi:2013ff,Barry:2013xxa,Doi:1985dx,Vergados:2002pv}
\begin{eqnarray}
  \left[T_{1/2}^{0\nu}\right]^{-1} &=& G^{0\nu}_{01}|{\cal M}^{0\nu}_\nu|^2|\eta_\nu + 
  \eta_S|^2. 
 \label{eq:halflife_simp}
\end{eqnarray}
where the dimensionless particle physics parameters are 
\begin{eqnarray}
\label{eqn:eta1}
& &\eta_{\nu} = \sum_{i} \frac{ (\mathcal{V}^{\nu \hat{\nu}}_{ei})^2\, 
m_i}{m_e},\quad \quad \eta_{S} 
= \sum_{i}\frac{ (\mathcal{V}^{\nu \hat{S}}_{ei})^2\, m_p}{M_{S_i}} 
\end{eqnarray}
In eqn.\,(\ref{eqn:eta1}), $m_e$ $(m_i)$= mass of electron 
(light neutrino), and $m_p$ = proton mass. In eqn.\,(\ref{eq:halflife_simp}), 
$G^{0\nu}_{01}$ is the the phase space factor and besides different 
particle parameters, it contains  the nuclear matrix elements due to 
different chiralities of the hadronic weak currents such as 
$\left(\mathcal{M}^{0\nu}_{\nu} \right)$ involving left-left 
chirality in the standard contribution. Explicit numerical values 
of these nuclear matrix elements discussed in ref.\cite{Doi:1985dx,
Vergados:2002pv,Barry:2013xxa,Awasthi:2013ff} are  given in 
Table.\,\ref{tab:nucl-matrix}.
\begin{table}[h]
\centering
\vspace{10pt}
\begin{tabular}{lcccc}
\hline \hline \\ 
\multirow{2}{*}{Isotope}  & $G^{0\nu}_{01}[10^{-14}{\rm yrs}^{-1}]$ & \multirow{2}{*}{${\cal M}^{0\nu}_\nu$} 
                                                         & \multirow{2}{*}{${\cal M}^{0\nu}_N$}   \\
            & Refs. \cite{Barry:2013xxa, Awasthi:2013ff} &  & \\
\hline \\
$^{76}$Ge   & 0.686  & 2.58--6.64  & 233--412  \\ 
$^{82}$Se   & 2.95   & 2.42--5.92  & 226--408  \\ 
$^{130}$Te  & 4.13   & 2.43--5.04  & 234--384  \\ 
$^{136}$Xe  & 4.24   & 1.57--3.85  & 160--172  \\ 
\hline \hline
\end{tabular}
\caption{Phase space factors and nuclear matrix elements with 
their allowed ranges as derived in Refs. 
\cite{Doi:1985dx,Vergados:2002pv,Barry:2013xxa,Awasthi:2013ff}.}
\label{tab:nucl-matrix}
\end{table} 

In terms of effective mass parameter, the inverse half-life for 
neutrinoless double beta decay is given as,
\begin{eqnarray}
\left[T^{\frac{1}{2}}_{0\nu} \right]^{-1} &= &\frac{\Gamma_{0\nu \beta \beta}}{\text{ln\,2}} = G_{0\nu}
\left|\frac{\mathcal{M}_{\nu}}{m_e}\right|^2 \times |m_{ee}^{\mathrm{eff}}|^2\, \, , \\
&\mbox{with}& m_{ee}^{\mathrm{eff}}= m^{\nu}_{ee} + m^{\rm S}_{ee} \, , \nonumber
\end{eqnarray}
where $G_{0\nu}$ contains the phase space factors, $m_e$ is the electron mass, and ${\mathcal{M}_{\nu}}$
is the nuclear matrix element and the effective mass parameters are 
\begin{eqnarray}
m_{\rm ee}^{\nu} = \mathcal{N}^2_{e\,i} m_{\nu_i}\quad, \quad  m_{\rm ee}^{S} 
=p^2\, \frac{\left(\mathcal{V}^{\nu \hat{S}}_{e\,i}\right)^2}{M_{S_i}} 
\, ,\label{eqn:mS-ee}
\end{eqnarray}
where $p^2=-|p^2|$. 
With $\big| \langle p^2\rangle \big|= \big|m_e m_p{ \mathcal{M}}^{0\nu}_{N}
/{\mathcal{M}}^{0\nu}_{\nu}\big| \simeq (120-200)\,\, \mbox{MeV}^2$,
$M_{W_R} \simeq 10^{5}$ GeV, we predict the effective mass for 
$0\nu \beta \beta$ transition rate for light neutrino masses,
\bea \label{eq:mee}
|m^{\nu}_{\rm ee}| =\mathcal{N}^2_{e\, 1}\, m_{\nu_1}+  \mathcal{N}^2_{e\, 2}\, m_{\nu_2} + 
            \mathcal{N}^2_{e\, 3}\, m_{\nu_3} \simeq \left\{
\baz 
0.004\, \mbox{eV}
& \mbox{ NH,} \\[0.2cm]
0.048\, \mbox{eV}
& \mbox{ IH,} \\[0.2cm]
0.23\, \mbox{eV} 
& \mbox{ QD.} 
\ea \right. 
\eea
\begin{figure}[h!]
\begin{center}
\includegraphics[scale=0.94]{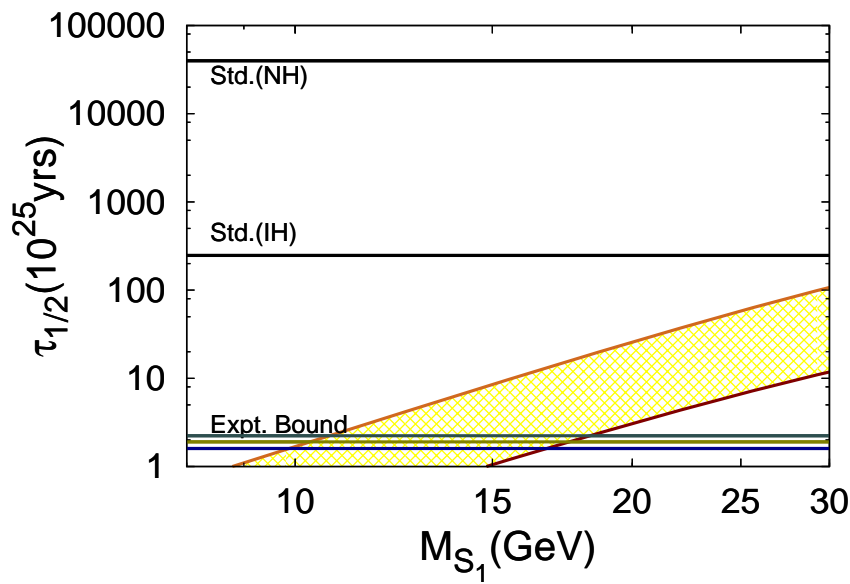}
\caption{Prediction of half life-time for neutrinoless double beta 
decay in this model in the $W^-_L - W^-_L$ channel as a function of lightest 
sterile neutrino mass $M_{S_1}$ for light NH neutrino masses (Yellow
band), but due to sterile neutrino exchanges. The band of uncertainty is due 
to the uncertainty in the neutrino virtuality momentum 
$|p|=120 {\rm MeV}-200 {\rm MeV}$. The upper dashed-horizontal 
lines are predictions only due to light neutrino exchanges of NH and 
IH patterns of masses. The lower horizontal lines are lower bounds of 
three experimental groups \cite{KlapdorKleingrothaus:2006bd,Auger:2012ar,Gando:2012zm,Agostini:2013mzu}.}    
\label{fig:lifetime}
\end{center}
\end{figure}

For direct prediction of half-life as a function of heavy sterile 
neutrino and its comparison with experimental data of ongoing 
search experiments, we derive the following analytic formula 
\begin{eqnarray}
T^{\frac{1}{2}}_{0\nu} &= &\mathcal{K}^{-1}_{0\nu} \times 
\frac{M^2_{N_1}\, M^4_{S_1}}{|\langle p^2\rangle |^2
\left(M_{D_{e1}}\right)^4} \bigg[\bigg| 1 + {\bf a}
\frac{M^2_{S_1}}{M^2_{S_2}} + {\bf b}  \frac{M^2_{S_1}}{M^2_{S_3}}-{\bm \delta} \bigg| 
\bigg]^{-2} \, ,\label{lifeformula}
\end{eqnarray}
where $K_{0\nu}=1.57 \times 10^{-25}\, \, \mbox{yrs}^{-1}\, \, \mbox{eV}^{-2}$ and 
\begin{eqnarray}
 {\bf a} =\frac{M_{D_{e2}}^2}{M_{D_{e1}}^2}\frac{M_{N_1}}{M_{N_2}}\quad, \quad 
  {\bf b}=\frac{M_{D_{e3}}^2}{M_{D_{e1}}^2}\frac{M_{N_1}}{M_{N_3}}\quad, \quad
  {\bm \delta}=\frac{m^{\nu}_{ee}M_{N_1}}{M^2_{D_{e1}}}\frac{M^2_{S_1}}{|p^2|}.\label{ab}
\end{eqnarray}
This formula is different from the one obtained using type-II seesaw
dominance in $SO(10)$ with TeV scale $Z^{prime}$ \cite{bidyut:2013}. 
Using the predicted value of $M_D$ from eq.(\ref{eq:md-mr0}) and 
derived values of heavy RH Majorana neutrino mass matrix,
$M_N=\mbox{diag}(115, 1750, 7500)$ GeV from the GUT-scale 
fit to the fermion masses we obtain from eq.(\ref{ab})
\be
{\bf a}  = 1.666-i\, 0.394\quad ,  \quad {\bf b} =-3.7815 
-i\,3.3456.
\ee

For different values of the diagonal matrix $M={\rm diag}(M_1, M_2, M_3)$
consistent within the non-unitarity constraint eq.(\ref{eqn:rel-eta}),
and the $M_N={\rm diag}(115, 1785, 7500)$~GeV, we derive mass eigenvalues 
$\hat{M}_S=(M_{S_1}, M_{S_2}, M_{S_3})$ using the formula
\be
\hat{M}_S=-(M^2_1/M_{N_1}, M^2_2/M_{N_2}, M^2_3/M_{N_3}).\label{eq:M_S_aprox}
\ee
The moduli of these eigenvalues and the corresponding elements of $M$ are
given in Table~{\ref{tab:ms_vs_m}}. It is clear from eq.(\ref{lifeformula})
that the half-life is a function of three mass eigenvalues $M_{S_1}$,
$M_{S_2}$ and $M_{S_3}$ while all other parameters are known.
For fixed values of $M_{S_2}=50$~GeV and $M_{S_3}=394$~GeV, we have 
plotted half-life against the lightest sterile neutrino mass $M_{S_1}$,
as shown in Fig.{\ref{fig:lifetime}} by neglecting the light neutrino
exchange contribution in eq.(\ref{lifeformula}) $({\bm \delta=0})$. 
\begin{table}
\begin{center}
\begin{tabular}[h!]{|l|l|l|}
\hline
$M$~(GeV)&$|{\hat{M}}_S|$~(GeV)[eq.(\protect\ref{eq:M_S_aprox})]& ${M_S}_{\rm exact}$~(GeV) [NH]  \\
\hline
(20, 300, 1718) & (3.478, 50.42, 393.5)& (3.379, 49.07, 376.5)\\
\hline
(25, 300, 1718)
&(5.435, 50.42, 393.5)& (5.199, 49.07, 376.5)\\
\hline
(30, 300, 1718)
&(7.826, 50.42, 393.5)& (7.354, 49.07, 376.5)\\
\hline
(35, 300, 1718)
&(10.65, 50.42, 393.5)& (9.812, 49.07, 376.5)\\
\hline
(40, 300, 1718)
&(13.91, 50.42, 393.5)& (12.54, 49.07, 376.5)\\
\hline
(45, 300, 1718)
&(17.61, 50.42, 393.5)& (15.51, 49.07, 376.5)\\
\hline
(50, 300, 1718)
&(21.74, 50.42, 393.5)& (18.69, 49.07, 376.5)\\
\hline
(55, 300, 1718)
&(26.30, 50.42, 393.5)& (22.06, 49.07, 376.5)\\
\hline
(60, 300, 1718)
&(31.30, 50.42, 393.5)& (25.59, 49.07, 376.5)\\
\hline
(65, 300, 1718)
&(36.74, 50.42, 393.5)& (29.27, 49.07, 376.5)\\
\hline
(70, 300, 1718)
&(42.61, 50.42, 393.5)& (33.07, 49.07, 376.5)\\
\hline
(75, 300, 1718)
&(48.91, 50.42, 393.5)& (36.99, 49.07, 376.5)\\
\hline
\end{tabular}
\caption{Eigenvalues of sterile neutrino mass matrix for different 
allowed $N-S$ mixing matrix elements.
}\label{tab:ms_vs_m}
\end{center}
\end{table} 

It is evident from eq.(\ref{lifeformula}) that for 
$M_{S_3}>> M_{S_2}>> M_{S_1}$, a $\log(T_{1/2})$ vs $\log(M_{S_1})$ 
would exhibit a linear behavior. 
The half-life for neutrinoless double beta decay is presented by the 
blue-color band of Fig.\ref{fig:lifetime} which is due to existing 
uncertainty in the nuclear matrix elements as well as the resulting 
range of allowed values of $|\langle p \rangle|=120$ MeV -$200$ MeV. 
The two dashed horizontal lines represent half-life predictions for 
the standard hierarchical (NH) and inverted hierarchical (IH) cases
by taking light neutrino exchange contribution only. The colored 
solid horizontal lines at the bottom of the figure represent 
recent experimental lower limits by three different groups 
\cite{KlapdorKleingrothaus:2006bd,Auger:2012ar,Gando:2012zm,Agostini:2013mzu}. 
It is quite clear from Fig.\ref{fig:lifetime} that the lightest of 
the three heavy sterile neutrino masses has the lower bound
\be
\left[M_{S_1} \right]_{\rm LNV}\geq 14\pm 4 {\rm GeV}.\label{lnvbound}
\ee
Further eq.(\ref{lifeformula}) also predicts that this lower 
bound would not be affected significantly as long as 
$M_{S_3}>> M_{S_2}>> M_{S_1}$ which is easily satisfied by the 
type of solutions allowed by the non-unitarity constraint on 
$\eta_{\alpha\beta}$. 
\begin{figure}
\begin{center}
\includegraphics[scale=.9]{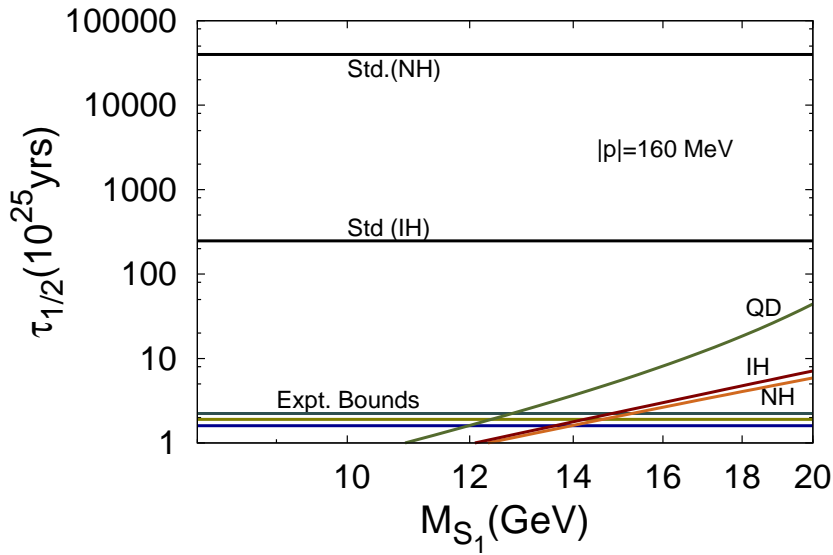}
\end{center}
\caption{Same as  Fig.\protect\ref{fig:lifetime} with sterile neutrino
  exchanges but for NH, 
IH, and QD patterns of light neutrino masses. The
value of neutrino virtuality momentum has been fixed at 
$|p| = 160$~MeV for all the three cases.
}\label{fig:nh_ih_qd_lifetime}
\end{figure}

Including the light neutrino exchange contribution in the formula for
NH, IH and QD cases $({\bm \delta}\neq 0)$ and for fixed value of $|p|=160$~MeV 
we have plotted the half-life as a function of $M_{S_1}$ as shown in 
Fig.{\ref{fig:nh_ih_qd_lifetime}} where the upper, lower, and the middle 
curves represent the QD, NH, and IH pattern of light neutrino masses, respectively.
Because of the opposite sign of the light-neutrino and the sterile-neutrino
exchange contributions, there is partial cancellation between the two 
 effective mass parameters especially in the QD case resulting 
in a somewhat larger life-time compared to the NH case or the IH case. This figure gives the
lower bound at $M_{S_1}\gtrsim 14$ GeV for NH and IH cases as before 
, but $M_{S_1}\gtrsim 12.5 $ GeV for the QD case. But when the whole
range of
uncertainty in the $|p^2|$ is included, this difference disappears.

We have checked that, although a knaive extrapolation of the formula of
eq.(\ref{lifeformula}) for higher eigen values of
$M_{S_i}(i=1,2,3)$ resulting from larger values of $M_1, M_2, M_3$
gives a maximum in the half-life finally settling down at $T_{1/2}
\sim 2\times10^{25}$ yrs., these solutions contradict the prediction
of TeV scale $Z^{\prime}$ boson accessible to LHC. They also correspond to
 LFV decay branching ratios  substantially lower
than those estimated here.

 It is clear that even when the
whole range of uncertainties in $|p|$ are included the three curves
overlap although for a fixed value of $|p|$ the QD case gives a factor
of nearly $7$ times larger lifetime prediction compared to the other
two cases especially when $M_{S_1}\simeq 20$\, GeV. This suggests that
even if the uncertainties in $|p|$, mainly emerging due to nuclear matrix
elements is considerably reduced, detection of
$0\nu\beta\beta$ decay life time in the QD case would require more
accurate measurements, if the light neutrinos are quasidegenerate and
$M_{S_1} = 15-20$ GeV. Because of the underlying gauged inverse nature
of the inverse seesaw mechanism relying upon the validity of the
constraints on the mass matrices $M_N\gg M \gg M_D,{\mu_S}$ and the
non-unitarity constraint emerging from LFV decays, large values of
$M_{S_1}$ beyond $20$ GeV are disallowed.    

Apart from these small and moderate cancellations between the two
contributions, we do not find any large cancellation leading to large
half-life inaccessible to ongoing $0\nu\beta\beta$ decay searches.   
We conclude  that the sterile neutrino exchange
contribution dominates the decay process in most of the regions of the
parameter space especially for NH and IH cases for all allowed
$M_{S_1}$ values. In the QD case this holds for $M_{S_1}$ values up to
$\sim 20$ GeV allowed by the model.

\section{\small SUMMARY AND DISCUSSIONS}\label{sec8}
We have implemented extended seesaw mechanism in a class of $SO(10)$ models
containing one additional fermion singlet ($S$) per generation leading 
to TeV scale $Z^{\prime}$ boson and heavy RH Majorana neutrino ($N$) 
masses via  $U(1)_R\times U(1)_{B-L}$ gauge symmetry breaking  generated 
through the Higgs representation ${126}_H$ while the $N-S$ mixing matrix 
$M$ is generated through the VEV of RH doublet Higgs contained in ${16}_H$. 
Inspite of the presence of the TeV scale RH neutrino mass matrix 
$M_N$, and naturally dominant Dirac neutrino mass matrix ($M_D$) 
in the model, the would-be large contribution due to type-I seesaw cancels out. 
The type-II seesaw contribution is damped out because of large parity 
violating scale and the TeV scale $B-L$ breaking. The formula for light left-handed 
neutrino masses and mixings are adequately well represented by the gauged 
inverse seesaw formula . The Dirac neutrino mass matrix $M_D$ that plays 
a crucial role in the inverse seesaw formula, non-unitarity effects and 
predictions of LFV decays and $0\nu\beta\beta$ decay 
is obtained by fitting the charged fermion masses and CKM mixings at the GUT 
scale for which the induced VEV of $\xi(2,2,15)\subset {126}_H$ is utilized 
in addition to two separate Higgs doublets originating from 
${10}_{H_{(1,2)}}$. The roles of two different types of $SO(10)$ 
structures corresponding  to the presence of (i) a single representation 
${126}_H$ leading to a diagonal structure of RH neutrino mass matrix,  or (ii) two representations ${126}_H$ and ${126}_H^{\prime}$ 
leading to a  general structure of RH neutrino mass matrix, are discussed with their respective impact on the 
phenomenology of observable $n-{\bar n}$ oscillation.  While the dominant
new contribution to  $0\nu 2 \beta$ decay  in the $W_L-W_L$ channel
due to sterile neutrino exchanges, saturates the current experimental
limits arrived at various experimental groups, the branching ratios for LFV decays, and rare 
kaon decays are noted to be within the accessible ranges of ongoing search 
experiments. Using RG analysis, we have derived the lower bound on the lepto-quark gage
boson mass mediating rare kaon decays to be $M_{lepto}\ge (1.53\pm
0.06)\times 10^6$ GeV which is easily accommodated in the GUT
scenario. The unification constraint on gauge couplings of the
$SO(10)$ model is found to permit diquark Higgs scalar masses
extending from $\sim$ (10-100) TeV leading to observable $n-{\bar n}$
oscillation while satisfying flavor physics constraints \cite{flavor-constraint}saturating the lepto-quark gauge boson mass
bound.  
These suggests that the model is also simultaneously consistent with
observable rarekaon decay by ongoing search experiments.

Compared to the recent interesting proposal of ref.\cite{Babu:2012vc,Babu:PRD,Babu:PRL}, 
although successful generation of baryon asymmetry of the universe has
not been implemented so far in this model, 
we have one extra gauge boson accessible to LHC. Likewise, 
in our model the lepto-quark gauge boson mediated $K_L\to \mu \bar{e}$ 
is also accessible to ongoing search experiments. Whereas the new $B-L$ 
violating proton decay is predicted to be accessible in ref.\cite{Babu:2012vc,Babu:PRD,Babu:PRL}, 
in our case it is $B-L$ conserving proton decay $p \to e^+\pi^0$. Whereas 
the type-I seesaw mechanism associated with high $B-L$ breaking 
scale is generally inaccessible to direct experimental tests, in 
our case the TeV-scale gauged inverse seesaw is directly verifiable. 
In the minimal model the predicted values of the RH neutrino masses 
are also accessible for verification at LHC. 

Even though the model is non-supersymmetric, it predicts similar 
branching ratios as in SUSY models for LFV processes like $\mu \to e\gamma$, 
$\tau \to \mu\gamma$, and $\tau \to e\gamma$. Even for the Dirac phase 
$\delta =0,\pi,2\pi$ of the PMNS matrix, the model predicts the
leptonic CP-violation 
parameter $J \simeq 10^{-5}$ due to non-unitarity effects. We have 
explicitly derived a new formula for the 
 half life of $0\nu\beta\beta$ 
decay as a function of the  sterile neutrino masses in the model 
and derived the lower bound $M_{S_1} \ge 14\pm 4$ GeV imposed 
by the current experimental limits on the half life. In this model as
also in the model of ref.\cite{Awasthi:2013ff}, 
the lifetime corresponding to Heidelberg Moscow experiment does not 
necessarily require the light neutrinos to be quasi-degenerate.
We have checked that the analytic formula obtained for the half life
for the dominant  $0\nu\beta\beta$ in the $W_L-W_L$ channel of
ref.\cite{Awasthi:2013ff}, is analogous to the formula obtained in this
work with almost the same lower bound on the lightest sterile
neutrino mass when the decay amplitudes in the RH sector is neglected
because of much heavier $W_R$ mass.

The predicted proton-lifetime in the minimal model is found to be 
${\tau}_p(p\to e^+\pi^0) \simeq 5.05 \times 10^{35\pm 1.0\pm 0.34}
{\rm yrs}$ where the first(second) uncertainty is due to GUT-threshold
effects(experimental errors). This lifetime is accessible to ongoing
and planned experiments. We have noted significant reduction of the predicted 
lifetime, bringing the central value much closer to the current Super K. limit with
${\tau}_p(p\to e^+\pi^0)=1.1\times 10^{34} {\rm yrs}-5.05\times
10^{35}$ yrs when the effect of a lighter bi-triplet Higgs contained in
the representation ${54}_H\subset SO(10)$ is included. 
 We conclude 
that even though the model does not have low-mass RH $W_R^{\pm}$ bosons 
in the accessible range of LHC, it is associated with interesting 
signatures on lepton flavor, lepton number and baryon number
violations and rare kaon decays.  
\newpage

\section{\small APPENDIX A}
\subsection{Estimation of experimental and GUT-threshold 
uncertainties on the unification scale}
\subsubsection{Analytic formulas}\label{subsub1app}
In contrast to other intermediate gauge symmetries, $SO(10)$ model 
with ${G}_{224D}$ intermediate symmetry was noted to have 
the remarkable property that GUT threshold corrections arising out 
of superheavy masses or higher dimensional operators identically 
vanish on $\sin^2\theta_W$ or the $G_{224D}$ breaking scale
\cite{parpat:1991,parpat:1992,mkp:1998,lmpr:1994vp}. We show how this 
property can be ensured in this model with precision gauge coupling 
unification while predicting vanishing GUT-threshold corrections on $M_P$, 
analytically, but with non-vanishing finite corrections on $M_{GUT}$. We 
derive the corresponding GUT threshold effects in $SO(10)$ model with 
three intermediate symmetry breaking steps, ${G}_{224D}$, 
${G}_{224}$, and ${G}_{2113}$ between the GUT 
and the standard model whereas the uncertainties in the mass 
scales has been discussed in ref.\cite{rnmmkp:1992} only with 
single intermediate breaking. The symmetry breaking chain under 
consideration is
\begin{eqnarray}
SO(10)\mathop{\longrightarrow}^{ {{\Large a^{\prime \prime \prime}_i}}}_{M_U}  {G}_{224D}
          \mathop{\longrightarrow}^{ {{\Large a^{\prime \prime}_i}}}_{M_P}
          {G}_{224} \mathop{\longrightarrow}^{ {{\Large a^\prime_i}}}_{M_C}
          {G}_{2113} \mathop{\longrightarrow}^{ {{\Large a_i}}}_{M^0_R}
          {G}_{\rm SM} \mathop{\longrightarrow}^{}_{M_Z} {G}_{13},
         \label{app:1}
\end{eqnarray} 
where ${\Large a^{\prime \prime \prime}_i}$, 
${\Large a^{\prime \prime}_i}$, ${\Large a^\prime_i}$, 
and ${\Large a_i}$ are, respectively, the one-loop beta 
coefficients for the gauge group ${G}_{2_L 2_R 4_C D}$, 
${G}_{2_L 2_R 4_C}$, ${G}_{2_L 1_{R} 1_{B-L} 3_C}$, 
and ${G}_{\rm SM}\equiv {G}_{2_L 1_Y 3_C}$.

Following the formalism used in ref.\cite{rnmmkp:1992, lmpr:1994vp}, 
one can write the expressions for two different contributions of 
$\sin^2 \theta_W \left(M_Z \right)$, and $\alpha_s \left(M_Z \right)$:
{\small 
\begin{eqnarray}
\hspace*{-0.55cm}16\pi\left(\alpha^{-1}_s -\frac{3}{8} 
\alpha^{-1}_{\rm em} \right)= \mathcal{A}_P 
\mbox{ln}\left( \frac{M_P}{M_Z}\right) +
\mathcal{A}_U \mbox{ln}\left( \frac{M_U}{M_Z}\right)
+\mathcal{A}_C \mbox{ln}\left( \frac{M_C}{M_Z}\right)+\mathcal{A}_0
\mbox{ln}\left( \frac{M^0_R}{M_Z}\right)+ f^U_M,
\label{app:2}
\end{eqnarray}
}
where,
\begin{eqnarray}
& &\hspace*{-0.4cm}\mathcal{A}_0=\left( 8 {a}_{3C} - 3 {a}_{2L} - 5{a}_Y \right)
- \left(8{a}^\prime_{3C} - 3 {a}^\prime_{2L} - 3 {a}^\prime_{1R} 
- 2 {a}^\prime_{B-L}\right)\,, \nonumber \\
& &\hspace*{-0.4cm}\mathcal{A}_C=\left(8{a}^\prime_{3C} 
- 3{a}^\prime_{2L} - 3 {a}^\prime_{1R} - 2 {a}^\prime_{B-L}\right)
- \left(6 {a}^{\prime \prime}_{4C} - 3{a}^{\prime \prime}_{2L} 
- 3{a}^{\prime \prime}_{2R}\right)\,, \nonumber \\
& &\hspace*{-0.4cm}\mathcal{A}_P=\left(6 {a}^{\prime \prime}_{4C} 
- 3{a}^{\prime \prime}_{2L} - 3{a}^{\prime \prime}_{2R}\right)
- \left( 6 {a}^{\prime \prime \prime}_{4C} 
- 6 {a}^{\prime \prime \prime}_{2L} \right)\,, \nonumber \\
& &\hspace*{-0.4cm}\mathcal{A}_U=\left( 6 {a}^{\prime \prime \prime}_{4C} 
- 6 {a}^{\prime \prime \prime}_{2L} \right)\, ,
                 \nonumber \\
& &f^U_M= \lambda^{U}_{2L}-\lambda^{U}_{4C} \, . \nonumber
\end{eqnarray}
Similarly,
{\small 
\begin{eqnarray}
\hspace*{-0.55cm} 16 \pi\, \alpha^{-1}_{\rm em} \left(\sin^2 \theta_W 
-\frac{3}{8} \right)= \mathcal{B}_P \mbox{ln}\left( \frac{M_P}{M_Z}\right) +
\mathcal{B}_U \mbox{ln}\left( \frac{M_U}{M_Z}\right) 
+ \mathcal{B}_C \mbox{ln}\left( \frac{M_C}{M_Z}\right)+\mathcal{B}_0
\mbox{ln}\left( \frac{M^0_R}{M_Z}\right)+ f^U_\theta,
\label{app:3}
\end{eqnarray}
}
with
\begin{eqnarray}
& &\mathcal{B}_0=\left( 5{a}_{2L} - 5{a}_{Y} \right)
           - \left( 5 {a}^{\prime}_{2L} - 3{a}^{\prime}_{1R} 
           - 2 {a}^{\prime}_{B-L} \right)\,, \nonumber \\
& &\mathcal{B}_C=\left( 5{a}^{\prime}_{2L} - 3{a}^{\prime}_{1R} 
- 2 {a}^{\prime}_{B-L} \right) - \left(5 {a}^{\prime \prime}_{2L} 
- 3{a}^{\prime \prime}_{2R} - 2{a}^{\prime \prime}_{4C} \right)\,, \nonumber \\
& &\mathcal{B}_P=\left(5 {a}^{\prime \prime}_{2L} 
- 3{a}^{\prime \prime}_{2R} - 2{a}^{\prime \prime}_{4C} \right)
- \left( 2{a}^{\prime \prime \prime}_{2L} 
- 2{a}^{\prime \prime \prime}_{4C} \right)\,,  \nonumber \\
& &\mathcal{B}_U=\left( 2{a}^{\prime \prime \prime}_{2L} 
- 2{a}^{\prime \prime \prime}_{4C} \right)\,,  \nonumber \\
& &f^U_\theta=\frac{1}{3} \left( \lambda^{U}_{4C}-\lambda^{U}_{2L}\right) \, . \nonumber
\end{eqnarray}

It is well known that threshold effects at intermediate scales 
are likely to introduce discontinuities in the gauge
couplings thereby destroying possibilities of precision unification. 
This fact has led us to restrict the model with
vanishing intermediate scale threshold corrections by assuming 
relevant sub-multiplets to have masses exactly equal
to their respective intermediate scales which is applicable to the 
intermediate scales $M_R^0$, $M_R^+$, and $M_C$
in the present work.

Denoting $\mathcal{C}_0=16 \pi \left(\alpha^{-1}_s 
-\frac{3}{8} \alpha^{-1}_{\rm em} \right)$, and $\mathcal{C}_1=
16 \pi\, \alpha^{-1}_{\rm em} \left(\sin^2 \theta_W -\frac{3}{8} \right)$, 
one can rewrite the eq.\,(\ref{app:2}), and eq.\,(\ref{app:3}) for $M_P$
and $M_U$ as
\begin{eqnarray}
& &\hspace*{-0.55cm}\mathcal{A}_P \mbox{ln}\left( \frac{M_P}{M_Z}\right) 
+ \mathcal{A}_U \mbox{ln}\left( \frac{M_U}{M_Z}\right) =\mathcal{D}_0
= \mathcal{C}_0 -\mathcal{A}_C \mbox{ln}\left( \frac{M_C}{M_Z}\right)
-\mathcal{A}_0 \mbox{ln}\left(\frac{M^0_R}{M_Z}\right) 
- f^U_M\, ,  \label{app:4} \\
& &\hspace*{-0.55cm}\mathcal{B}_P \mbox{ln}\left( \frac{M_P}{M_Z}\right) 
+ \mathcal{B}_U \mbox{ln}\left( \frac{M_U}{M_Z}\right) =\mathcal{D}_1
= \mathcal{C}_1 - \mathcal{B}_C \mbox{ln}\left( \frac{M_C}{M_Z}\right) 
- \mathcal{B}_0 \mbox{ln}\left(\frac{M^0_R}{M_Z}\right) - f^U_\theta\, . 
\nonumber \\  \label{app:5}
\end{eqnarray}
A formal solution for these two sets of eqns.\,(\ref{app:4}), and (\ref{app:5}),
\begin{eqnarray}
& &\mbox{ln}\left( \frac{M_U}{M_Z}\right)=\frac{\mathcal{D}_1 \mathcal{A}_P-\mathcal{D}_0 \mathcal{B}_P}
                {\mathcal{B}_U \mathcal{A}_P-\mathcal{A}_U \mathcal{B}_P}\, ,
                \label{app:6} \\
& &\mbox{ln}\left( \frac{M_P}{M_Z}\right)=\frac{\mathcal{D}_0 \mathcal{B}_U-\mathcal{D}_1 \mathcal{A}_U}
                {\mathcal{B}_U \mathcal{A}_P-\mathcal{A}_U \mathcal{B}_P} \, .  
                \label{app:7}
\end{eqnarray}
In this present work, we derive two types of uncertainties in 
the mass scales of $SO(10)$ model; i.e, the first one comes from 
low energy parameters taken from their experimental errors and 
another one arising from the threshold corrections accounting 
the theoretical uncertainties in the mass scales due to heavy
Higgs fields present at GUT scale. These two categories are 
presented below:

\noindent
\subsubsection{Uncertainties due to experimental 
errors in $\sin^2 \theta_W$ and $\alpha_s$}\label{subsub2app}

In eqns.\,(\ref{app:4}) and (\ref{app:5}) the low energy 
parameters are contained in $\mathcal{C}_0$ and $\mathcal{C}_1$.
As a result, we have got further simplified relations relevant 
for experimental uncertainties, i.e, $\Delta \left(\mathcal{D}_0\right)
= \Delta \left(\mathcal{C}_0\right)$ and $\Delta \left(\mathcal{D}_1\right)
= \Delta \left(\mathcal{C}_1\right)$, and hence,
\begin{eqnarray}
\Delta\, \mbox{ln}\left(\frac{M_U}{M_Z} \right)\,\bigg|_{\rm expt.}& &=
                \frac{\Delta \left( \mathcal{C}_1\right) \mathcal{A}_P- \Delta \left( \mathcal{C}_0\right) \mathcal{B}_P}
                {\mathcal{B}_U \mathcal{A}_P-\mathcal{A}_U \mathcal{B}_P} \nonumber \\
           &&=  \frac{\left[(16\pi)\, \alpha^{-1}_{\rm em} (\delta \sin^2\theta_W)\right]  \mathcal{A}_P
               -\left[-\frac{(16 \pi)}{\alpha^2_s}\left(\delta \alpha_s\right)\right] \mathcal{B}_P}
                {\mathcal{B}_U \mathcal{A}_P-\mathcal{A}_U \mathcal{B}_P}\, , \\
\Delta\, \mbox{ln}\left(\frac{M_P}{M_Z} \right)\,\bigg|_{\rm expt.}& &=
                \frac{\Delta \left( \mathcal{C}_0\right) \mathcal{B}_U- \Delta \left( \mathcal{C}_1\right) \mathcal{A}_U}
                {\mathcal{B}_U \mathcal{A}_P-\mathcal{A}_U \mathcal{B}_P}\nonumber \\
         && =  \frac{\left[-\frac{(16 \pi)}{\alpha^2_s}\left(\delta \alpha_s\right)\right] \mathcal{B}_U
               - \left[(16\pi)\, \alpha^{-1}_{\rm em} (\delta \sin^2\theta_W)\right] \mathcal{A}_U}
                {\mathcal{B}_U \mathcal{A}_P-\mathcal{A}_U \mathcal{B}_P}      \, ,
\end{eqnarray}
where, the errors in the experimental values on electroweak 
mixing angle $\sin^2 \theta_W$ and strong coupling
constant $\alpha_s$ as $\sin^2 \theta_W = 0.23102 \mp 0.00005, 
\quad \alpha_s =0.118 \pm 0.003$ giving $\delta \alpha_s=\pm 0.003$ 
and $\delta \sin^2\theta_W=\mp 0.00005$.

\noindent
\subsubsection{ Uncertainties in $M_U$ with vanishing correction on $M_P$}\label{subsub3app}

In the present work, we have considered minimal set of Higgs 
fields belonging to a larger $SO(10)$ Higgs representation 
implying other Higgs fields which do not take part in symmetry 
breaking will automatically present at GUT scale. Since we can 
not determine the masses of these heavy Higgs bosons and, hence, they
introduce uncertainty in other mass scales $M_P$ and $M_U$ via 
renormalization group equations resulting source of GUT threshold 
uncertainty in our predictions for proton life time. For this 
particular model, the GUT threshold corrections to D-parity breaking 
scale and unification mass scale is presented below
\begin{eqnarray}
\Delta \mbox{ln}\left(\frac{M_U}{M_Z} \right)\bigg|_{\mbox{\small GUT~ Th.}}
          &&      =\frac{\Delta \left( \mathcal{D}_1 \right) \mathcal{A}_P- \Delta \left( \mathcal{D}_0 \right) \mathcal{B}_P}
                {\mathcal{B}_U \mathcal{A}_P-\mathcal{A}_U \mathcal{B}_P} \nonumber \\
          &&=        \frac{- f^U_M}{6\left({a}^{\prime \prime \prime}_{2L} - {a}^{\prime \prime \prime}_{4C}\right)}\, , \\
\Delta \mbox{ln}\left(\frac{M_P}{M_Z} \right)\bigg|_{\mbox{\small GUT~ Th.}}
          &&      = \frac{ \Delta \left( \mathcal{D}_0 \right) \mathcal{B}_U- \Delta \left(\mathcal{D}_1 \right) \mathcal{A}_U}
                {\mathcal{B}_U \mathcal{A}_P-\mathcal{A}_U \mathcal{B}_P}\nonumber \\
          &&=         \frac{\mathcal{B}_U\, f^U_M - \mathcal{A}_U\, f^U_\theta}
                {24\,\left({a}^{\prime \prime \prime}_{2L} - {a}^{\prime \prime \prime}_{4C} \right)
                \left({a}^{\prime \prime}_{2L} - {a}^{\prime \prime}_{4C} \right)} =\mbox{0}\, .
\end{eqnarray}

\begin{center}
\hspace*{2cm}
\begin{table}
\begin{tabular}{|c|c|c|c|}
\hline
&&& \\
Group $G_{I}$        & Higgs content    & $ { a_i}$       & ${ b_{ij}}$            \\
\hline \hline
&&&\\
${\small G_{1_Y2_L3_C}}$                & 
$\begin{array}{l}
\Phi(\frac{1}{2},2,1)_{10}\end{array}$
                               & 
$\bmt 
41/10 \\
-19/6 \\
-7
\emt$ 
                               &  
$\bmt 
199/50   & 27/10    & 44/5  \\
9/10     & 35/6     & 12    \\
11/10    & 9/2      & -26
\emt $ \\
\hline 
&&&\\
${\small G_{1_{B-L}1_R 2_L 3_C}}$                                                                  &  
${\small \begin{array}{l}
\Phi_1(0,\frac{1}{2},2,1)_{10}\oplus \Phi_2(0,-\frac{1}{2},2,1)_{10^\prime}\\[2mm]
\Delta_R(-1,1,1,1)_{126} \oplus \chi_R(-\frac{1}{2},\frac{1}{2},1,1)_{16} \end{array}}$   
                                                                                          &
$\bmt 
37/8 \\ 
57/12\\
-3\\
-7
\emt$                                                                                     
                                                                                          &
$\bmt
209/16   & 63/8   & 9/4  & 4 \\
63/8     & 33/4   & 3    & 12\\
3/2      & 1      & 8    & 12\\
1/2      & 3/2    & 9/2  & -26
\emt$ \\
\hline 
&&&\\
${\small G_{2_L2_R4_C}}$                          & 
${\small \begin{array}{l}
\Phi_1(2,2,1)_{10}\oplus \Phi_2(2,2,1)_{10^\prime}\\[2mm]
\Delta_R(1,3,\overline{10})_{126} \oplus \chi_R(1,2,\overline{4})_{16}\\[2mm]
\sigma_R(1,3,15)_{210} 
\end{array}}$
                                         &
$\bmt -8/3 \\
       29/3\\
      -16/3 \emt$                        &
$\bmt 
37/3    & 6       & 45/2   \\
6       & 1103/3  & 1275/2 \\
9/2     & 255/2   & 736/3
\emt$\\
\hline 
&&&\\
${\small G_{2_L 2_R 4_C D}}$                                              &   
${\small \begin{array}{l}
\Phi_1(2,2,1)_{10}\oplus \Phi_2(2,2,1)_{10^\prime}\\[2mm]
\Delta_L(3,1,10)_{126} \oplus \Delta_R(1,3,\overline{10})_{126}\\[2mm]
\chi_L(2,1,4)_{16} \oplus \chi_R(1,2,\overline{4})_{16}\\[2mm]
\sigma_L(3,1,15)_{210} \oplus \sigma_R(1,3,15)_{210} \\[2mm]
\xi (2,2,15)_{126/126^\prime}
  \end{array}}$
                                                                  &
$\bmt 44/3 \\
      44/3 \\
      16/3 \emt$                                                  &
$\bmt 
1298/3  & 51      & 1755/2 \\
51       & 1298/3  & 1755/2 \\
351/2   & 351/2   & 1403/2
\emt$\\
\hline
\end{tabular}
\caption{One and two loop beta coefficients for different 
gauge coupling evolutions described in text taking the second 
Higgs doublet at $\mu \geq 5$\,TeV.}
\label{tab:beta_coeff}
\end{table}
\end{center}

The last step resulting in vanishing GUT-threshod correction
analytically follows by using expressions for $f_M^U,f_{\theta}^U$
,${\mathcal {B}_U}$ and $\mathcal {A}_U$ derived in
\ref{subsub1app}. This was proved in ref.\cite{mkp:1998}.

\newpage \noindent
{\bf ACKNOWLEDGEMENT}

\noindent Ram Lal Awasthi acknowledges  hospitality at the Center 
of Excellence in Theoretical and Mathematical Sciences, SOA University 
where this work was initiated and completed. M. K. P. thanks the
Department of Science and Technology, Government of India for a
research project.



\begin{thebibliography}{}
\bibitem{Dirac:1925}
P.\,A.\,M.\, Dirac, 
  ``{\em {The Fundamental Equations for Quantum Mechanics}}'',
    \href{}{Proceedings of the Royal Society of London\, {\bf 109} (1925) 752,\,pp. 642-653}.

\bibitem{Majorana:1937vz}
E. Majorana, 
  ``{\em {Theory of the Symmetry of Electrons and Positrons}}'',
    \href{}{Nuovo Cim.\, {\bf 14} (1937) 171-184}.

\bibitem{KlapdorKleingrothaus:2000sn}
H.\,V.\, Klapdor-Kleingrothaus, A.\, Dietz, L.\, Baudis, G.\, Heusser, I.\,V.\, Krivosheina, S.\, Kolb, 
B.\, Majorovits, H.\, Pas {\em et al.}, 
   ``{\em {Latest Results from the Heidelberg-Moscow Double Beta Decay Experiment}}'',
     \href{http://dx.doi.org/10.1007/s100500170022}{Eur. Phys. J. {\bf A\,12} (2001) 147-154},
     \href{http://arxiv.org/abs/hep-ph/0103062}{arXiv:0103062 [hep-ph]}. 
     
\bibitem{Aalseth:2002rf}
{\bf IGEX Collaboration}, C.\,E.,\,Aalseth,  
   ``{\em {The IGEX Ge-76 neutrinoless double beta decay experiment: Prospects for next generation experiments}}'',
     \href{http://dx.doi.org/10.1103/PhysRevD.65.092007}{Phys.\, Rev.\, {\bf D\,65} (2002) 092007}.
     \href{http://arxiv.org/abs/0202026 }{arXiv:0202026 [hep-ex]}.
     
\bibitem{KlapdorKleingrothaus:2004wj}
H.V. Klapdor-Kleingrothaus, I.V. Krivosheina, A. Dietz, and O. Chkvorets, 
   ``{\em {Search for neutrinoless double beta decay with enriched $^{76}$Ge in Gran Sasso 1990-2003}}'',
     \href{http://dx.doi.org/10.1016/j.physletb.2004.02.025}{Phys.\,Lett.\, {\bf B\,586} (2004) 198-212}.
     \href{http://arxiv.org/abs/0404088}{arXiv:0404088 [hep-ph]}.

\bibitem{KlapdorKleingrothaus:2006bd}
H.V. Klapdor-Kleingrothaus, I.V. Krivosheina, and I.V. Titkova, 
   ``{\em {Theoretical investigation of pulse shapes of double beta events in a $^{76}$Ge detector, 
           their dependence on particle physics parameters, and their separability from background 
           gamma events}}'',
     \href{http://dx.doi.org/10.1142/S0217732306020524}{Mod.\,Phys.\,Lett.\, {\bf A\,21} (2006) 1257-1278 }.
  
\bibitem{Gando:2012zm}
{\bf KamLAND-Zen Collaboration}, A.~Gando {\em et al.}, 
  ``{\em {Limit on Neutrinoless $\beta\beta$ Decay of $^{136}$Xe from the First Phase of 
    KamLAND-Zen and Comparison with the Positive Claim in $^{76}$Ge}}'',
    \href{http://dx.doi.org/10.1103/PhysRevLett.110.062502}{Phys.\,Rev.\, Lett.\, {\bf 110} (2013)  062502},
    \href{http://arxiv.org/abs/1211.3863}{{\tt arXiv:1211.3863 [hep-ex]}}.
    
\bibitem{Auger:2012ar}
{\bf EXO Collaboration}, M.\,Auger {\em et al.}, 
  ``{\em {Search for Neutrinoless Double-Beta Decay in $^{136}$Xe with EXO-200}}'',
    \href{http://dx.doi.org/10.1103/PhysRevLett.109.032505}{Phys.\,Rev.\, Lett.\, {\bf 109} (2012) 032505}.
    \href{http://arxiv.org/abs/1205.5608 }{arXiv:1205.5608 [hep-ex]}.
    
\bibitem{Agostini:2013mzu}
{\bf GERDA Collaboration}, M.\,Agostini {\em et al.}, 
  ``{\em {Results on neutrinoless double beta decay of $^{76}$Ge from GERDA Phase I}}'',
    \href{http://dx.doi.org/10.1103/PhysRevLett.111.122503}{Phys.\,Rev.\,Lett.\, {\bf 111} (2013) 122503}.
    \href{http://arxiv.org/abs/1307.4720 }{arXiv:1307.4720 [nucl-ex]}.

  
\bibitem{Arnaboldi:2008ds}
{\bf CUORICINO Collaboration}, C. Arnaboldi {\em et al.}, 
   ``{\em {Results from a search for the 0 neutrino beta beta-decay of $^{130}$Te}}'',
   \href{http://dx.doi.org/10.1103/PhysRevC.78.035502}{Phys.\,Rev.\,{\bf C\,78 } (2008) 035502}.
   \href{http://arxiv.org/abs/0802.3439 }{arXiv:0802.3439 [hep-ex]}.
   
\bibitem{Minkowski:1977sc}
P.\,Minkowski, 
  ``{\em {$\mu \to e \gamma$ at a Rate of One Out of 1-Billion Muon Decays?}}'',
    \href{http://dx.doi.org/}{Phys.\,Lett.\, {\bf B\,67} (1977) 421}.

\bibitem{Yanagida:1979as}
T.\,Yanagida, ``{\em {Horizontal gauge symmetry and masses of neutrinos}}'',
  \href{http://dx.doi.org/}{In Proceedings of the Workshop on the Baryon Number of the Universe 
                and Unified Theories, Tsukuba, Japan, 13-14 Feb. 1979}.

\bibitem{Mohapatra:1979ia}
R.\,N.\, Mohapatra, and G.\,Senjanovic,
  ``{\em {Neutrino mass and spontaneous parity nonconservation}}'',
    \href{http://dx.doi.org/10.1103/PhysRevLett.44.912}{Phys.\,Rev.\,Lett.\, {\bf 44} (1980) 912}.

\bibitem{Schechter:1980gr}
J.\, Schechter and J.\,W.\,F.\, Valle, 
  ``{\em {Neutrino Masses in $SU(2) \times U(1)$ Theories}}'',
    \href{http://dx.doi.org/10.1103/PhysRevD.22.2227}{Phys.\,Rev.\,{\bf D\,22} (1980) 2227}.

\bibitem{Magg:1980ut}
M.\,Magg and C.\,Wetterich, 
  ``{\em {Neutrino Mass Problem and Gauge Hierarchy}}'',
    \href{http://dx.doi.org/10.1016/0370-2693(80)90825-4}{Phys.\,Lett.\, {\bf B\,94} (1980) 61}.

\bibitem{Lazarides:1980nt}
G.~Lazarides, Q.~Shafi, and C.~Wetterich, 
  ``{\em {Proton Lifetime and Fermion Masses in an SO(10) Model}}'',
    \href{http://dx.doi.org/10.1016/0550-3213(81)90354-0}{Nucl.Phys. {\bf B\,181} (1981) 287}.
  

\bibitem{Ibarra:2010xw}
A.~Ibarra, E.~Molinaro, and S.~Petcov, ``{\em {TeV Scale See-Saw Mechanisms of
  Neutrino Mass Generation, the Majorana Nature of the Heavy Singlet Neutrinos
  and $(\beta\beta)_{0\nu}$-Decay}}'',
  \href{http://dx.doi.org/10.1007/JHEP09(2010)108}{JHEP {\bf 1009} (2010)
  108},
  \href{http://arxiv.org/abs/1007.2378}{{\tt arXiv:1007.2378 [hep-ph]}}.

%
%
%

\bibitem{Mohapatra:1974gc}
R.~N,~Mohapatra and J.~C. Pati, 
  ``{\em {A Natural Left-Right Symmetry}}'',
    \href{http://dx.doi.org/10.1103/PhysRevD.11.2558}{Phys.Rev. {\bf D\,11}, 2558 (1975)}.

\bibitem{Pati:1974yy}
J.~C. Pati and A.~Salam, 
  ``{\em {Lepton Number as the Fourth Color}}'',
    \href{http://dx.doi.org/10.1103/PhysRevD.10.275,10.1103/PhysRevD.11.703}{Phys. Rev. {\bf D\,10}, 275 (1974)}.

\bibitem{Senjanovic:1975rk}
G.~Senjanovic and R.~N. Mohapatra, 
  ``{\em {Exact Left-Right Symmetry and Spontaneous Violation of Parity}}'',
    \href{http://dx.doi.org/10.1103/PhysRevD.12.1502}{Phys. Rev. {\bf D\,12},1502 (1975)}.

\bibitem{Mohapatra:1980yp}
R.~N. Mohapatra and G.~Senjanovic, 
  ``{\em {Neutrino Masses and Mixings in Gauge Models with Spontaneous Parity Violation}}'',
    \href{http://dx.doi.org/10.1103/PhysRevD.23.165}{Phys.Rev. {\bf D23} (1981) 165}.
    
  
\bibitem{Mohapatra:1980qe}
R.~N.~Mohapatra and R.~E.~Marshak, 
  ``{\em {Local B-L Symmetry of Electroweak Interactions, Majorana Neutrinos and Neutron Oscillations}}'',
    \href{http://dx.doi.org/10.1103/PhysRevLett.44.1316}{Phys.\, Rev.\, Lett.\, {\bf 44} (1980) 1316-1319}.
\bibitem{Moha-Senj:1982}
R.~N.~Mohapatra, G.~Senjanovic,
``{\em {Higgs-boson effects in grand unified theories}}'',
\href{http://dx.doi.org/10.1103/PhysRevD.27.1601}{Phys. Rev. .{\bf D 27}\, (1983) 1601.}
\bibitem{Mohapatra:JPG} For a recent review see R. N. Mohapatra,
  J. Phys. {\bf G. 36}\,(2009)104006.  
\bibitem{Aguila:1981}
F.~Del~Aguilla, L.~Ibanez,
``{\em{Higgs bosons in SO(10) and partial unification}}'',
\href{http://dx.doi.org/10.1016/0550-3213(81)90266-2}{Nucl. Phys. {\bf
    B 177}\,(1981) 60}.
         
  

\bibitem{Babu:2012vc}
K.~S.~Babu and R.~N.~Mohapatra, 
  ``{\em {Coupling Unification, GUT-Scale Baryogenesis and Neutron-Antineutron Oscillation in SO(10) }}'',
    \href{http://dx.doi.org/10.1016/j.physletb.2012.08.006}{Phys.\,Lett.\, {\bf B\,715} (2012) 328-334}.
      \href{http://arxiv.org/abs/1206.5701}{{\tt arXiv:1206.5701
          [hep-ph]}}
\bibitem{Babu:PRD}K. S. Babu and R. N. Mohapatra, ``{\em{B-L violating
  nucleon decay and GUT-scale baryogenesis in
  $SO(10)$}}'',{Phys. Rev. {\bf D 86} (2012) 035018}.

\bibitem{Babu:PRL}K. S. Babu and R. N. Mohapatra, ``{\em{B-L violating
  proton decay modes and new baryogenesis scenario in
  $SO(10)$}}'',{Phys. Rev. Lett. {\bf 109} (2012) 091803}.
\bibitem{ATLAS:extrawz}ATLAS Collaboration, G. Ad {\em et al.},
  Eur. Phys. J. {\bf C 72} (2012) 2056 [arxiv:1203.5420].
\bibitem{CMS:extrawz}CMS Collaboration, Phys. Rev. Lett.{\bf 109} (2012)261802[arxiv:1210.2402]. 
\bibitem{Chang:1983fu}
D.~Chang, R.~N.~Mohapatra and M.~K.~Parida,  
  ``{\em {Decoupling Parity and $SU(2)_R$ Breaking Scales: A New Approach to Left-Right Symmetric Models}}'',
    \href{http://dx.doi.org/10.1103/PhysRevLett.52.1072 }{Phys.\,Rev.\,Lett.\, {\bf 52} (1984) 1072}.

\bibitem{Chang:1984wp}
D.~Chang, R.~N.~Mohapatra, J.~Gipson,R.~E.~Marshak and M.~K.~Parida, 
  ``{\em {Experimental Tests of New $SO(10)$ Grand Unification}}'',
    \href{http://dx.doi.org/10.1103/PhysRevD.31.1718 }{Phys.\,Rev.\, {\bf D\,31} (1985) 1718}.
      
\bibitem{mkp83}M. K. Parida, ``{\em {Natural mass scales for matter
    anti-matter oscillations in SO(10)}}''
\href{http://dx.doi.org/10.1016/0370-2693(83)90594-4}{Phys. Lett. {\bf
    B 126}\,(1983) 220}.  
\bibitem{mkpPRD83}M. K. Parida,''{\em Matter-anti-matter oscillations
  in grand unified theories with high unification masses}'', \href{http://dx.doi.org/10.1103/PhysRevD.27.2783}
{Phys. Rev. {\bf D  27}\,(1983) 2783}.

\bibitem{dambrose98}D. Ambrose {\em et al.}\,[BNL
  Collaboration],\,''{\em{New limit on electron and muon number
    violation from $K_L \to \mu^{\pm}e^{\mp}$ decay}}'',
\href{http://dx.doi.org/10.1163/PhysRevLett.81.5734}{Phys. Rev. Lett. {\bf
    81} \,(1998) 5734;[hep-ex/9811038]}.

      
\bibitem{Awasthi:2013ff}
Ram.~L. Awasthi, M.~K.~Parida, and Sudhanwa ~Patra, ``{\em {Neutrino masses, dominant
  neutrinoless double beta decay, and observable lepton flavor violation in
  left-right models and SO(10) grand unification with low mass $ W_R, Z_R$
  bosons}}'',
\href{http://dx.doi.org/10.1007/JHEP08(2013)122}{JHEP\, {\bf 08} (2013) 122}.
\href{http://arxiv.org/abs/1302.0672}{arXiv:1302.0672 [hep-ph]}.
\bibitem{zprime-th}P. Langacker, Rev. Mod. Phys. {\bf 81} (2009) 1199;
T. Han, P. Langacker, Z. Liu, L. -T. Wang, arXiv:1308.2738;
   M. K. Parida and A. Raychaudhuri, Phys. Rev. {\bf D 26}, (1982)
  2364; M. K. Parida, C. C. Hazra, Phys. Lett. {\bf B 121}, (1983)
  355; G. J. P. Eboli, J. Gonzalez-Fraile, M. Gonzalez-Gartia,
  arxiv:1112.0316[hep-ph]; A. Falkowski, C. Grojean, A. Kaminska,
  S. Pokorski, and A. Wailer, JHEP {\bf 1111}(2011)028. 
\bibitem{Weinberg:1980}S. Weinberg, Phys. Lett. {\bf B 91}\,(1980)51.
\bibitem{Hall:1981} L. Hall, Nucl. Phys. {\bf B 178}\,(1981)75.
\bibitem{Ovrut:1981} B. Ovrut and H. Schnitzer {\bf B 196}\,(1981)163.
\bibitem{parpat:1991}
M.~K.~Parida and P.~K.~Patra, 
  ``{\em {Useful theorem on vanishing threshold contribution to $sin^2_{\theta_W}$ in a class of grand unified theories }}'',
    \href{http://dx.doi.org/10.1103/PhysRevLett.66.858 }{Phys.\,Rev.\,Lett.\, {\bf 66} (1991) 858-861}.

\bibitem{parpat:1992}
M.~K.~Parida and P.~K.~Patra, 
``{\em {Theorem on vanishing multiloop radiative corrections to 
$\sim^2_{\theta_W}$ in grand unified theories at high mass scales}}'', 
\href{http://dx.doi.org/10.1103/PhysRevLett.68.754}{Phys.\,Rev.\,Lett.\, {\bf 68} (1992) 754}.
\bibitem{rnmPLB:1993} R. N. Mohapatra,''{\em A Theorem on threshold
  corrections in grand unified theories
}'',{Phys. Lett. {\bf B 285}\,(1992) 235 }.
\bibitem{Langacker:1993} P. Langacker,\,N. Polonsky,''{\em
  Uncertainties in coupling unification}'',{Phys. Rev. {\bf D 47}\,(1993)4028}.
\bibitem{mkp:1998}
M.~K.~Parida, ``{\em {Vanishing corrections on the intermediate scale and 
implication for unification of forces}}'',
\href{http://dx.doi.org/10.1103/PhysRevD.57.2736}{Phys.\,Rev.\,{\bf D\, 57} (1998) 2736}[hep-ph/9710246].

\bibitem{rnmmkp:1992}
R.~N.~Mohapatra and M.~K.~Parida, 
  ``{\em {Threshold effects on the mass scale predictions in $SO(10)$ models and solar neutrino puzzle }}'',
    \href{http://dx.doi.org/10.1103/PhysRevD.47.264 }{Phys.\,Rev.\, {\bf D\,47} (1993) 264}.
      \href{http://arxiv.org/abs/9204234}{{\tt arXiv:9204234 [hep-ph]}}.
    
\bibitem{lmpr:1994vp}
Dae-Gyu Lee, R.~N.~ Mohapatra, M.~K.~ Parida and M.~ Rani, 
  ``{\em {Predictions for proton lifetime in minimal nonsupersymmetric $SO(10)$ models: An update }}'',
    \href{http://dx.doi.org/10.1103/PhysRevD.51.229 }{Phys.\,Rev.\, {\bf D\,51} (1991) 229}.
    \href{http://arxiv.org/abs/9404238}{{\tt arXiv:9404238 [hep-ph]}}. 
\bibitem{mkp:other-th} M. K. Parida and C. C. Hazra, Phys. Rev. {\bf D
  40}, 3074 (1989); M. K. Parida and M. Rani, Phys. Rev. {\bf D 49},
  (1994) 3704;
  M. K. Parida, Phys. Lett. {\bf B 196} (1987) 163; M. K. Parida,
  and B. D. Cajee, Eur. Phys. J. {\bf C 44}  (2005) 447; S. K. Majee,
  M. K. Parida, A. Raychaudhuri, U. Sarkar, Phys. Rev. {\bf D 75}, (2007)075003.
    
\bibitem{Yao:2006}
{\bf Particle Data Group}, W.-M. Yao {\em et al.}. 
  ``{\em {Particle Data Group: partial update for edition 2008 (URL: http://pdg.lbl.gov)}}'', 
    \href{http://dx.doi.org/J. Phys. {\bf G\,33}, 1 (2006)}{J. Phys. {\bf G\,33}, 1 (2006)}.
      
\bibitem{PhysRevLett.56.561}
R.~N.~Mohapatra, 
  ``{\em {Mechanism for understanding small neutrino mass in superstring theories }}'',
    \href{http://dx.doi.org/10.1103/PhysRevLett.56.561}{Phys.~Rev.~Lett.\, {\bf 56} (1986) 561}.
      
\bibitem{PhysRevD.34.1642}
R.~N.~Mohapatra and J.~W.~F.~ Valle, 
  ``{\em { Neutrino mass and baryon-number nonconservation in superstring models}}'',
    \href{http://dx.doi.org/10.1103/PhysRevD.34.1642 }{Phys.\,Rev.\, {\bf D\,34} (1986) 1642}.

 \bibitem{Wyler:1982dd}
D.~Wyler and L.~Wolfenstein,
  ``{\em {Massless Neutrinos in Left-Right Symmetric Models}}'',
    \href{http://dx.doi.org/10.1016/0550-3213(83)90482-0 }{Nucl.\,Phys.\,{\bf B\,218} (1983) 205}.

\bibitem{mkpARC:2010}
M.~K.~Parida and A.~Raychaudhuri, ``Inverse seesaw, leptogenesis, observable proton decay
and $\Delta^{\pm\pm}_R$ in SUSY SO(10) with heavy $W_R$'', 
{Phys.\,Rev.\,{\bf D\, 82} (2010) 093017}. 


   
\bibitem{Majee:2008mn}
S.~K.~ Majee, M.~K.~ Parida and A.~ Raychaudhuri, 
  ``{\em {Neutrino mass and low-scale leptogenesis in a testable SUSY $SO(10)$ model }}'',
    \href{http://dx.doi.org/10.1016/j.physletb.2008.08.048 }{Phys.\,Lett.\, {\bf B\,668} (2008) 299}.
      \href{http://arxiv.org/abs/0807.3959}{{\tt arXiv:0807.3959 [hep-ph]}}.   
      
\bibitem{ap:2011aa}
Ram L. Awasthi and M.~K.~Parida, 
  ``{\em {Inverse Seesaw Mechanism in Nonsupersymmetric $SO(10)$, Proton Lifetime, 
  Nonunitarity Effects, and a Low-mass $Z^\prime$ Boson }}'', 
    \href{http://dx.doi.org/10.1103/PhysRevD.86.093004 }{Phys.\,Rev. {\bf D\,86} (2012) 093004}.
      \href{http://arxiv.org/abs/}{{\tt arXiv: [hep-ph]}}.
\bibitem{fuku:2002} T. Fukuyama and T. Kikuchi,''{\em {Renormalisaton group
  equations for quark lepton mass matrices in the SO(10) model with
  Higgs scalars}}'',
\href{http://arxiv.org/10.1142/S021773009848}
{Mod. Phys. Lett. {\bf A 18}\,(2003) 719}[hep-ph/0206118].
\bibitem{Grimus:2000vj}
W.~Grimus and L.~Lavoura, 
  ``{\em {The Seesaw mechanism at arbitrary order: Disentangling the small scale from the large scale}}'', 
    \href{http://dx.doi.org/10.1016/}{JHEP {\bf 0011} (2000)  042}, 
    \href{http://arxiv.org/abs/hep-ph/0008179}{{\tt arXiv:hep-ph/0008179 [hep-ph]}}.
      
      

\bibitem{Fogli} G. Fogli, E. Lisi, A. Marrone, A. Palazzo, and
  A. Rotunno, Phys. Rev. {\bf D 84}, (2011) 053007 [arxiv:1106.6028];
  T. Schwetz, M. Tortola and J. W. F. Valle, New J. Phys. {\bf 13},
  (2011)063004 [arxiv: 1103.0734]; D. Forero, M. Tortola, and
  J. W. F. Valle, Phys. Rev. {\bf D 86}, (2012) 073012 [arxiv:1205.4018].  
\bibitem{Ilakovac:1994kj}
A.~Ilakovac and A.~Pilaftsis, 
  ``{\em {Flavor violating charged lepton decays in seesaw-type models}}'', 
    \href{http://dx.doi.org/10.1016/0550-3213(94)00567-X}{Nucl.\,Phys.\, {\bf B\,437} (1995) 491}, 
    \href{http://arxiv.org/abs/hep-ph/9403398}{{\tt arXiv:hep-ph/9403398 [hep-ph]}}. 
    
\bibitem{Cirigliano:2004mv}
V.~Cirigliano, A.~Kurylov, M.~Ramsey-Musolf, and P.~Vogel, 
  ``{\em {Lepton flavor violation without supersymmetry}}'',
    \href{http://dx.doi.org/10.1103/PhysRevD.70.075007}{Phys.\,Rev.\, {\bf D\,70} (2004) 075007}, 
    \href{http://arxiv.org/abs/hep-ph/0404233}{{\tt arXiv:hep-ph/0404233 [hep-ph]}}.

\bibitem{Leontaris:1985qc}
G.~Leontaris, K.~Tamvakis, and J.~Vergados, 
  ``{\em {Lepton and family number violation from exotic scalars}}'', 
    \href{http://dx.doi.org/10.1016/0370-2693(85)91078-0}{Phys.\,Lett.\, {\bf B\,162} (1985) 153}.  

\bibitem{Ilakovac:2012sh}
A.~Ilakovac, A.~Pilaftsis, and L.~Popov, 
   ``{\em {Charged Lepton Flavour Violation in Supersymmetric Low-Scale Seesaw Models}}'',
     \href{http://arxiv.org/abs/1212.5939}{{\tt arXiv:1212.5939 [hep-ph]}}. 
      
\bibitem{Adam:2013mnn} 
{\bf MEG Collaboration}, J.~Adam {\em et al.}, 
  ``{\em {New constraint on the existence of the $\mu^+ \to e^+ \gamma$ decay}}'', 
    \href{http://arxiv.org/abs/1303.0754}{{\tt arXiv:1303.0754 [hep-ex]}}.


%
%
%
%
%
%
%
\bibitem{Barry:2013xxa}
J.~Barry and W.~Rodejohann, 
  ``{\em {Lepton number and flavour violation in TeV-scale left-right symmetric theories with large left-right mixing }}'', 
    \href{http://dx.doi.org/10.1007/JHEP09(2013)153 }{JHEP\, {\bf 1309} (2013) 153}.
%
       
\bibitem{Doi:1985dx}
M.~Doi, T.~Kotani, and E.~Takasugi, 
  ``{\em {Double beta Decay and Majorana Neutrino}}'', 
    \href{http://dx.doi.org/10.1143/PTPS.83.1}{Prog.Theor.Phys.Suppl. {\bf 83} (1985) 1}. 
  
\bibitem{Vergados:2002pv}
J.~Vergados, 
  ``{\em {The Neutrinoless double beta decay from a modern perspective}}'', 
    \href{http://dx.doi.org/10.1016/S0370-1573(01)00068-0}{Phys.Rept. {\bf 361} (2002) 1-56}, 
    \href{http://arxiv.org/abs/hep-ph/0209347}{{\tt arXiv:hep-ph/0209347 [hep-ph]}}.  

\bibitem{Babu:1992ia}
K.~S.~ Babu and R.~N.~ Mohapatra, 
  ``{\em {Predictive neutrino spectrum in minimal $SO(10)$ grand unification }}'',
    \href{http://dx.doi.org/10.1103/PhysRevLett.70.2845}{Phys.\,Rev.\,Lett.\, {\bf 70} (1993) 2845}.
    \href{http://arxiv.org/abs/9209215}{{\tt arXiv:9209215 [hep-ph]}}.

\bibitem{CMS:diquark} S. Chatrchyan {em et al.} (CMS Collaboration),
  JHEP {\bf 1301} (2013)013 arxiv:1210.2387[hep-ex]. 

\bibitem{Pnath:2007} P. Nath, P. F. Perez, Phys. Rep. {\bf 441}\,
  (2007)191.  

\bibitem{Parida:2011wh}
M.~K.~ Parida, 
  ``{\em {Radiative Seesaw in $SO(10)$ with Dark Matter }}'',
    \href{http://dx.doi.org/10.1016/j.physletb.2011.09.016}{Phys.\,Lett.\, {\bf B\,704} (2011) 206}.
      \href{http://arxiv.org/abs/1106.4137}{{\tt arXiv:1106.4137 [hep-ph]}}. 


\bibitem{Bertolini:2013vta}
S.~ Bertolini, L.~ Di Luzio and M.~Malinsky,  
  ``{\em {Light color octet scalars in the minimal $SO(10)$ grand unification }}'',
    \href{http://dx.doi.org/ }{Phys.\,Rev. {\bf D\,87} (2013) 085020}.
      \href{http://arxiv.org/abs/1302.3401}{{\tt arXiv:1302.3401
          [hep-ph]}}.



\bibitem{Nishino:2012ipa}
{\bf Super-Kamiokande Collaboration}, H.~ Nishino {\em et al.}
  ``{\em {Search for Nucleon Decay into Charged Anti-lepton plus Meson in Super-Kamiokande I and II }}'',
    \href{http://dx.doi.org/10.1103/PhysRevD.85.112001}{Phys.\,Rev. {\bf D\,85} (2012) 112001}.
      \href{http://arxiv.org/abs/1203.4030}{{\tt arXiv:1203.4030 [hep-ph]}}.
    
\bibitem{Parida:1996xa}  
M.~K.~ Parida and B. Purkayastha
  ``{\em {New lower bound on $SU(4)_C$ gauge boson mass from CERN-LEP measurements and $K_{L} \to \mu e$}}'', 
    \href{http://dx.doi.org/10.1103/PhysRevD.53.1706 }{Phys.\,Rev.\, {\bf D\,53} (1996) 1706}.
  
\bibitem{Deshpande:1982bf}
N. G. Deshpande and R. J. Johnson, 
  ``{\em {Experimental limit on $SU(4)_C$ gauge boson mass}}'',
    \href{http://dx.doi.org/10.1103/PhysRevD.27.1193 }{Phys.\,Rev.\, {\bf D\,27} (1984)  1193 }.
  
\bibitem{Arisaka:1992xy}
K.~ Arisaka {\em et al}, 
  ``{\em {Improved upper limit on the branching ratio $\mbox{Br} (K^0_L \to mu^\pm e^\mp)$ }}'',
    \href{http://dx.doi.org/10.1103/PhysRevLett.70.1049 }{Phys.\,Rev.\,Lett.\, {\bf 70} (1993) 1049}.   
    
    
\bibitem{nnbar-osciln-expt}
K.\, Genezer, 
  ``{\em {In Proceedings of the Workshop on (B-L) Violation, Lawrence Berkeley Laboratory, 2007 
          [http://inpa.lbl.gov/BLNV/blnv.htm]}}  
\bibitem{Keung:1983uu}
W.-Y. Keung and G.~Senjanovi{\'c}, ``{\em {Majorana neutrinos and the
  production of the right-handed charged gauge boson}}'',
\href{http://dx.doi.org/10.1103/PhysRevLett.50.1427}{Phys. Rev. Lett. {\bf 50}
  (1983)  1427}..  
\bibitem{jraaf:2012} 
J. L.\,Raaf[Super-Kamiokande Collaboration], ``{\em {Recent nucleon
    decay results from Super-Kamiokande}}'', 
\href{http://dx.doi.org/10.1016/j.nuclphysbps.2012.09.196}{Nucl. Phys. Proc. Suppl.{\bf
    229-232}\,(2012) 559}.
\bibitem{Hewett:2012} 
J. L. Hewett {\em et al.}, ``{\em {Fundamental physics at the
    intensity frontier}}'',
      {\tt arXiv: 1205.2671[hep-ex]}.
\bibitem{babuetal:2013} 
K. S. Babu {\em et al.}, ``{\em  {Baryon number violation}}'',
      {\tt arXiv:1311.5285 [hep-ph]}.  
  
\bibitem{dimo:1981} S. Dimopoulos, S. Raby, and G. L. Kane, Nucl. Phys.{\bf 182}, 77 (1981); 

\bibitem{dp:2001} C. R. Das, M. K. Parida,''{\em{New formulas and
    predictions of running fermion masses at higher scales in SM,
    2HDM, and MSSM}}'',
\href{http://arxiv.org/10.1007/s100520100628}{Eur. Phys. J. {\bf C 20} (2001) 121 [hep-ph/0010004]}.
\bibitem{antusch} S. Antusch, J. P. Baumann, and E. Fernandez-Martinez,
  Nucl. Phys. {\bf B\,810}, 369 (2009); S. Antusch, M. Blennow,
  E. Fernandez-Martinez, and J. Lopez-Pavon, Phys. Rev. {\bf D\,80}, 033002 (2009);
  S. Antush,
  C. Biggio, E. Fernandez-Martinez, M. Belen Gavela, and J. Lopez-Pavon, J. High
  Energy Phys. {\bf 10} (2006) 084; D. V. Forero, S. Morisi, M. Tartola and
  J. W. F. Valle, J. High Energy Phys. {\bf 09} (2011) 142.
\bibitem{non-unit} E. Fernandez-Martinez, M. B. Gavela, J. Lopez-Pavon and O. Yasuda, Phys. Lett. {\bf B\,649}, 427 (2007);  
                   K. Kanaya, Prog. Theor. Phys.,{\bf 64},2278 (1980); 
                   J. Kersten and A. Y. Smirnov, Phys. Rev. {\bf D\,76}, 073005 (2007); 
                   M. Malinsky, T. Ohlsson, H. Zhang, Phys. Rev. {\bf D\, 79}, 073009 (2009); 
                   G. Altarelli and D. Meloni, Nucl. Phys. {\bf B\, 809}, 158 (2009); 
                   F. del Aguila and J. A. Aguilar-Saavedra, Phys. Lett. {\bf B\, 672}, 158 (2009); 
                   F. del Aguila and J. A. Aguilar-Saavedra and J. de Blas, Acta Phys. Polon. {\bf B\,40}, 
                   2901 (2009); arXiv:0910.2720 [hep-ph]; 
                   A. van der Schaaf, J. Phys. {\bf G\, 29}, 2755
                   (2003).
\bibitem{bidyut:2013}B. P. Nayak and M. K. Parida, arxiv:2013.3185[hep-ph]. 
\bibitem{flavor-constraint} E. C. F. S. Fortes, K. S. Babu and
  R. N. Mohapatra, arxiv:1311.4101[hep-ph]
 

\end{thebibliography}
\end{document}